# Combining the Morris Method and Multiple Error Metrics to Assess Aquifer Characteristics and Recharge in the Lower Ticino Basin, in Italy

Authors:
Emily A. Baker[a,c], Alessandro Cappato[a], Sara Todeschini[a], Lorenzo Tamellini[b], Giancarlo Sangalli[b,c], Alessandro Reali[a,b], Sauro Manenti[a]

Affiliations:
[a]Department of Civil Engineering and Architecture (DICAr) and Research Centre on Water (CRA), University of Pavia, Via Ferrata, 3, 27100, Pavia, Italy
[b] Consiglio Nazionale delle Ricerche - Istituto di Matematica Applicata e Tecnologie Informatiche "E. Magenes" (CNR-IMATI), Via Ferrata 5 / A, 27100, Pavia, Italy
[c]Department of Mathematics, University of Pavia, Via Ferrata, 5, 27100, Pavia, Italy

Corresponding Author: Emily A. Baker
e-mail: emily.baker@unipv.it

## Highlights
- The Morris Method ranks the most influential uncertain parameters on model outputs.
- Uninfluential parameters can be identified and set constant in subsequent analyses.
- Using multiple error metrics provides more realistic estimates of input parameters.
- Rice field irrigation is a major source (~88%) of late summer Ticino basin recharge.
- In Pavia, ~3.3% of streamflow is due to groundwater inflow from Vigevano to Pavia.

## Abstract
Groundwater flow model accuracy is often limited by the uncertainty in model parameters that characterize aquifer properties and aquifer recharge. Aquifer properties such as hydraulic conductivity can have an uncertainty spanning orders of magnitude. Meanwhile, parameters used to configure model boundary conditions can introduce additional uncertainty. In this study, the Morris Method sensitivity analysis is performed on multiple quantities of interest to assess the sensitivity of a steady-state groundwater flow model to uncertain input parameters. The Morris Method determines which of these parameters are less influential on model outputs. Uninfluential parameters can be set constant during subsequent parameter optimization to reduce computational expense. Combining multiple quantities of interest (e.g., RMSE, groundwater fluxes) when performing both the Morris Method and parameter optimization offers a more complete assessment of groundwater models, providing a more reliable and physically consistent estimate of uncertain parameters. The parameter optimization procedure also provides us an estimate of the residual uncertainty in the parameter values, resulting in a more complete estimate of the remaining uncertainty. By employing such techniques, the current study was able to estimate the aquifer hydraulic conductivity and recharge rate due to rice field irrigation in a groundwater basin in Northern Italy, revealing that a significant proportion of surficial aquifer recharge (approximately 81-94%) during the later summer is due to the flood irrigation practices applied to these fields.

# 1. Introduction

Groundwater flow models are often associated with large uncertainties due to a lack of subsurface data and the heterogeneity of aquifer materials, leading to large uncertainties in the values of model input parameters such as hydraulic conductivity, specific yield and/or specific storage, and model boundary conditions. Due to the large number of uncertain input parameters, it can often be difficult and computationally demanding to determine which set of parameter values most accurately represents a groundwater system. Such uncertainties can result in the inability to make useful predictions from the groundwater flow model, such as failing to constrain an informative range in the modeled groundwater residence times. The Morris Method (Morris, 1991) is a global sensitivity analysis used to assess the sensitivity of model outputs to input parameters and can be used to reduce the uncertainty associated with groundwater flow models (Bianchi Janetti et al., 2019), allowing for greater insights into the behavior of the groundwater system. By assessing each input parameter's contribution to model uncertainty, parameters that do not affect model outputs can be set constant and therefore be excluded from further exploration (Kazakis, 2018; Bianchi Janetti et al., 2019). By eliminating less influential parameters, more effort can be exerted on exploring influential input parameters. In addition, the Morris Method is less computationally expensive than similar types of sensitivity analyses (Campolongo et al., 2007; Ruano et al., 2012; Nguyen & Reiter, 2015). Such a technique is useful in constraining groundwater models with numerous uncertain input parameters. While the Morris Method can be used to reduce the number of uncertain input parameters, defining a set of model parameters that are physically consistent and reliably reproduce the groundwater flow can still be difficult for domains with complex geometries and where multiple uncertain factors interact. For this reason, the present study implements a modeling approach that applies the Morris Method to multiple quantities of interest derived from the model simulations, combined with a joint error metric optimization to determine an optimal set of input parameters together with an estimate on their residual uncertainty.

This study applies the Morris Method to help constrain the input parameters to a steady-state groundwater flow model of the lower half of the Ticino basin within the Po Plain of Northern Italy. The Po Plain is an important agricultural area, responsible for most of the rice production in Italy, the largest producer of rice in Europe (Facchi et al., 2018; Zampieri et al., 2019). Within the lower half of the Ticino basin, almost one-third of the land surface is used for rice production (Regione Lombardia, 2019), which can have a large impact on the groundwater system due to the use of flood irrigation practices. Flooded rice fields in the Po Plain contain a continuous ponded water depth of about 5 – 10 cm (Cesari de Maria et al., 2016; Cesari de Maria et al., 2017). Over the entire growing season, the average total irrigation depth in fields that use flood irrigation can range from 1500 to 3000 mm but can vary depending on the year and due to soil characteristics and groundwater depth (Cesari de Maria et al., 2016; Cesari de Maria et al., 2017). Other types of rice cultivation techniques can reduce the irrigation requirement by 20-60%, decreasing the total irrigation depth for the season to about 600-1200 mm (Cesari de Maria et al., 2017). Flood irrigation can result in infiltration rates of 2.4 to 15.3 mm/day (Facchi et al., 2018), with some rice fields experiencing average infiltration rates of 10 to 40 mm/day with a standard deviation of 6 to 20 mm/day (Cesari de Maria et al., 2017). In some areas of the Po Plain, flood irrigation activities are known to raise groundwater levels up to 4-5 m during the irrigation season (Rotiroti et al., 2019; Lasagna et al., 2020). In addition to raising groundwater levels, the irrigation waters, which are sourced predominantly from the river, have been shown to help dilute high nitrate concentrations in the groundwater (Rotiroti et al., 2019).

While efforts have been made in some areas to change the types of irrigation practices applied to rice fields to reduce water use, this may not be the best approach given the concern that climate change will lead to less rainfall and higher temperatures in the region, resulting in a decrease in groundwater levels (Lasagna et al., 2020). Disruption to the artificial recharge of the aquifer through surface irrigation, which has been employed in the region since the 12$^{th}$ century, by employing more efficient irrigation methods could decrease aquifer recharge, resulting in lower groundwater levels and higher nitrate concentrations in the groundwater. Such changes would affect the quality and quantity of groundwater available at extraction wells. In addition, decreased groundwater levels would results in less flow to springs which sustain local micro-environments and provide smaller local sources of irrigation waters (Rotiroti et al., 2019; Lasagna et al., 2020). While the irrigation applied to rice fields is clearly an important recharge component affecting the hydrologic cycle in the Po Plain, the recharge rate due to irrigation is difficult to estimate a priori over such a large area. For this reason, the current work aims at developing a model to accurately represent the groundwater flow patterns in the lower Ticino basin through proper definition of aquifer properties and a reliable estimate of irrigation recharge rate.

In this study, a steady-state groundwater flow model is developed for the lower Ticino basin, extending approximately from the city of Abbiategrasso (upstream) to Pavia (downstream), to gain a better understanding of the aquifer properties (e.g., hydraulic conductivity) and estimate the amount of aquifer recharge within the lower Ticino basin due to flood irrigation at the end of the irrigation season. The model is calibrated for the period August/September 2014, when observed groundwater head data are available. The calibration is performed using a Negative Log Likelihood function, which combines the root mean squared error of observed groundwater heads with the percentage of grid cells with calculated heads above the land surface; this percentage serves as a proxy for the locations of springs within the study area that are unmapped or differ slightly in location due to model simplifications. While previous studies have assessed the recharge contributions from flood irrigation or developed groundwater flow models in portions of the Po Plain, those studies have not focused on the Ticino basin, but rather portions of the Po Plain to the west within the region of Piedmont (Lasagna et al., 2020) or further east in the region of Lombardy around Milan (De Caro et al., 2020) and the Adda (Balestrini et al., 2021; Musacchio et al., 2021) and Oglio (Rotiroti et al., 2019) river basins (Vassena et al., 2012). Other studies have assessed infiltration rates within rice fields in the Ticino basin, but only in a few sets of experimental fields (Cesari de Maria et al., 2016; Cesari de Maria et al., 2017: Facchi et al., 2018) rather than assessing the magnitude of recharge from flood irrigation techniques over a larger area of the basin, such as that one considered in this work. By coupling the Morris Method with multiple quantities of interest and using a combined error metric in subsequent parameter optimization, a more reliable estimate of the model input parameters, including the recharge due to rice field irrigation, is obtained, using fewer computational resources.

## 2. Materials & Methods
### 2.1 Study Area
The study area is in the lower portion of the Ticino hydrogeologic basin, within the Po Plain in the Lombardy region of Northern Italy (Fig. 1a). The 501.5 km$^2$ study area comprises the lower 41% of the Ticino basin and stretches about 42.6 km in the north-south direction from north of the town of Abbiategrasso to south of the city of Pavia and is typically about 12-14 km wide. The northern boundary of the study area is located just north of Abbiategrasso near where the aquifer units within the basin transition from being partially connected/weakly confined to distinct aquifer units consisting of a surficial unconfined aquifer underlain by confined units that may locally be semi-confined due to leakage through the aquitard (Musacchio et al., 2021). The land surface elevation in the study area ranges from about 53-124 m above sea level (asl), with an abrupt change in surface elevation of about 20-30 m from the higher elevation plains along the western and eastern edges to the lower elevation central river valley (Fig. 1c). The Ticino river is the main river in the study area, flowing south and slightly east through the approximate center of the basin until its confluence with the Po river along the southern border of the study area. The study area contains both natural (risorgive) and human enhanced (fontanili) springs where groundwater rises to the surface (Fig. 1c; Regione Lombardia, 2007a; Regione Lombardia, 2013b; De Luca et al., 2014; Balestrini et al., 2021).

Land usage in the considered portion of the Ticino basin is predominantly agricultural. Rice fields constitute 29.0% of the land surface in the study area (Regione Lombardia, 2019). Another 35.0% of the study area consists of non-irrigated arable land (27.4%) and other types of agricultural fields (7.6%). The remainder of the land in the study area is divided between urban/residential areas (15.8%) and natural environments such as forests, shrubland, wetlands, and dune/beach areas predominantly found along the Ticino river (Regione Lombardia, 2019). Rice fields are flooded starting in mid-April to early May and remain flooded until the end of August or September, when they are permitted to dry naturally (Lasagna et al., 2020; Balestrini et al., 2021). This irrigation technique contributes to the recharge of the surficial aquifer, with an estimated 40-50% or more of the applied water becoming aquifer recharge (Regione Lombardia, 2008; Lasagna et al., 2020). An extensive network of predominantly unlined canals and irrigation ditches, whose flow is frequently controlled and/or diverted, also traverse the basin (Fig. 1c), further contributing to aquifer recharge due to leakage (Pilla et al., 2006; Vassena et al., 2012; Pognant et al., 2013; Clemente et al., 2015). These canals are used to provide irrigation water to the rice fields. The water in these canals is a combination of water from the Ticino river and from local fontanili, but flowrate information is very poor. Therefore, the unconfined surface aquifer is recharged by both precipitation and irrigation activities, with flood irrigation common throughout the region due to the widespread cultivation of rice (Lasagna et al., 2020).

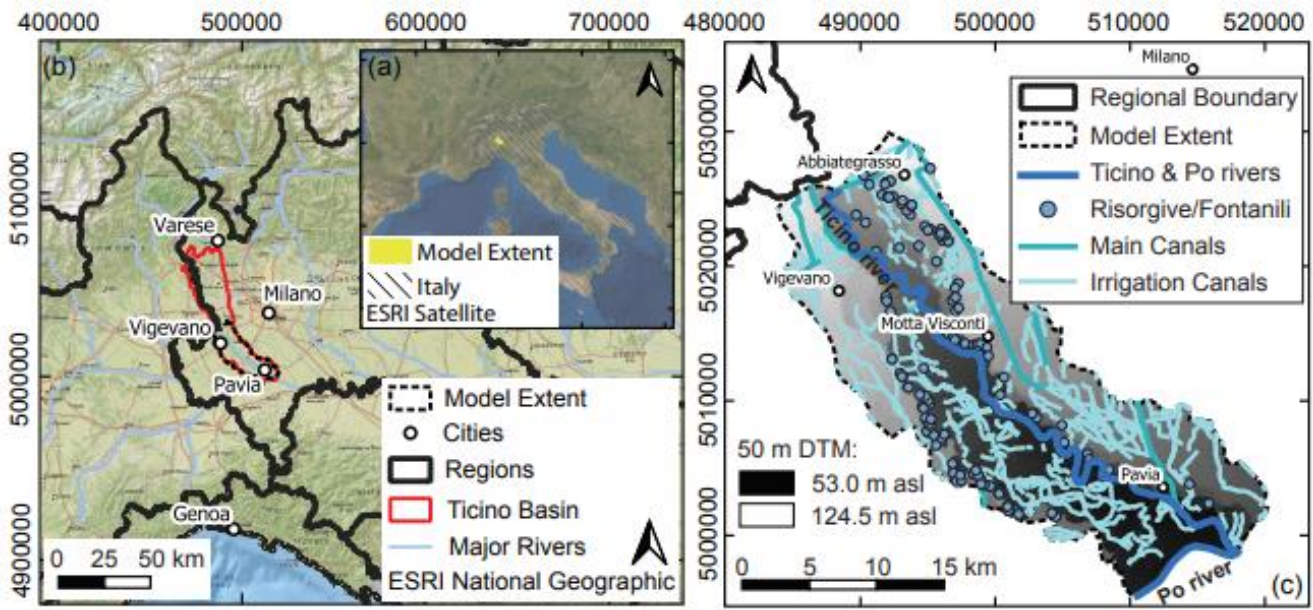

**Figure 1**. Map of the study region. (a) Location of the field site within Northern Italy. (b) Location of the study area within the region of Lombardy in the southern half of the approximately 1203 km$^2$ Ticino basin. (c) Hydrologic boundaries including rivers, canals, and mapped springs (risorgive & fontanili) within the study area. The DTM shows the land surface elevation in meters above sea level within the model extent.

The Ticino basin is underlain by multiple aquifer units. The surficial unconfined aquifer (termed the group A aquifer) consists of proximal braid plain deposits made up of gravel with a sandy matrix that was deposited during the middle-late Pleistocene (De Caro et al., 2020). Previous studies in the Lombardy region estimate the unconfined layer typically ranges from about 20 - 100 m thick (De Caro et al., 2020), with thicknesses of 80 – 130 m or more along the Ticino basin (Pilla et al., 2006; Bonomi et al., 2010). For this study, data on the elevation of the unconfined aquifer basal elevations were interpolated to create the basal surface of the model domain (Regione Lombardia, 2022). In the study area, the surficial aquifer has a range in thickness from about 35-109 m (mean of 64.2 m). The thinnest areas are in the river valley and the thickest portions are in the lateral plains along the basin edges and in the northern portion of the study area near Abbiategrasso (Fig. 2b). The hydraulic conductivity ($K$) of the unconfined aquifer in the Ticino basin decreases from north to south (De Caro et al., 2020). Previous studies in the Po Plain and adjacent Piedmont Plain indicate the hydraulic conductivity ranges from about $2 \times 10^{-4} - 1 \times 10^{-3}$ m/s, with values as low as $1 \times 10^{-5} - 1 \times 10^{-4}$ m/s and as high as $2 \times 10^{-3} - 5 \times 10^{-3}$ m/s (De Caro et al., 2020; Lasagna et al., 2020). The underlying group B aquifer is separated from the surficial unconfined aquifer by a clayey silty aquitard layer. This aquitard is continuous in the southern portion of the Ticino basin and disappears in the northern portion of the basin north of the study area. Additional deeper confined aquifers also occur below the group B aquifer (Pilla et al., 2006; Vassena et al., 2012; De Caro et al., 2020). In this study, only the surficial (group A) aquifer is modeled, and a no-flow boundary is located at the base of the model at the elevation of the aquitard.

**2.2 Numerical Model**

This study uses the modular hydrogeologic numerical model version 6 (MODFLOW 6) to simulate the groundwater heads and fluxes in the study area (Harbaugh, 2005; Langevin et al., 2017; Langevin et al., 2021). MODFLOW is an open-source code for groundwater flow modeling and simulation, developed by the United States Geological Survey (USGS). MODFLOW calculates the distribution of groundwater heads and flows through space and/or time, adopting a control-volume finite-difference discretization of the local mass-balance equation coupled with Darcy's law for groundwater flow:

$$\frac{\partial}{\partial x}\left(K_{xx}\frac{\partial h}{\partial x}\right) + \frac{\partial}{\partial y}\left(K_{yy}\frac{\partial h}{\partial y}\right) + \frac{\partial}{\partial z}\left(K_{zz}\frac{\partial h}{\partial z}\right) + W = S_s \frac{\partial h}{\partial t} \qquad [1]$$

where:
- $K_{xx}$, $K_{yy}$, and $K_{zz}$ are the principal components of the hydraulic conductivity tensor (L/T) along the coordinate directions $x$, $y$, and $z$ (vertical direction);
- $h$ is the piezometric head (L);
- $W$ is a volumetric flux per unit volume representing sources ($W>0$) and/or sinks ($W<0$) of water in the system ($T^{-1}$);
- $S_S$ is the specific storage of the porous material ($L^{-1}$);
- $t$ is time (T); (Harbaugh, 2005).

The control volumes are the model cells which are used for the discretization of both the aquifer material and governing Eq. 1 that occurs in each cell, with $h$ representing the value of the hydraulic head at each cell center (Langevin et al., 2017). This results in a coupled system of equations which are solved by assuming the total flow into and out of each cell equals the change in the rate of storage for each cell (Harbaugh, 2005). This system of equations is then iteratively solved within each cell using the initial conditions provided by the user. MODFLOW 6 is the most recent version of MODFLOW that aims to merge the capabilities of previous MODFLOW versions (Langevin et al., 2017). For this study, MODFLOW 6 was run through the FloPy package, which is a set of scripts designed by the USGS to run MODFLOW models through Python (Bakker et al., 2016).

The model of an unconfined aquifer leads to a nonlinear problem from the mathematical viewpoint due to the free-boundary, that is, the unknown interface between the saturated and unsaturated regions.

This study uses the Newton-Raphson method (the formulation previously implemented in MODFLOW-NWT; Hunt & Feinstein, 2012) rather than the standard formulation (previously MODFLOW-2005; Harbaugh, 2005) to solve the groundwater flow equation, because the former results in better model stability and improves convergence, especially in model scenarios where model cells dry and rewet, or when the water table traverses more than one cell layer due to complex geology (Niswonger et al., 2011; Langevin et al., 2017). During preliminary model runs, the Newton-Raphson method improved model stability since this method is more capable of dealing with the steep terrain occurring at the transition between the relatively high Ticino plain and the topographically lower Ticino river valley.

The mathematical problem is complemented by suitable boundary conditions. Multiple packages exist within the MODFLOW program that allow the user to define model boundary conditions and specify flux rates into and out of the model (Harbaugh, 2005). Model boundary conditions including the rivers (CHD package) and study area boundaries (GHD package) are parameterized using measured data interpolated to each grid cell, while the aquifer recharge rate (RCH package) due to precipitation is directly estimated by calculating the runoff and evapotranspiration and subtracting these losses from the precipitation. Since the groundwater model does not consider the unsaturated zone, the RCH boundary condition approximates the flux from the unsaturated zone to the saturated zone. Springs and fontanili are incorporated as drain boundaries (DRN package). Then the hydraulic conductivity values of the aquifer and an additional recharge rate (RCH package) due to flood irrigation within the rice fields and leakage from the irrigation canals are estimated through model calibration.

**2.2.1 Model Structure**

This work focuses on the steady-state groundwater flow model of the southern half of the Ticino river basin. The model domain is bounded to the east and west by the edges of the Ticino groundwater basin. The groundwater model consists of three layers of grid cells, representing the unconfined surficial aquifer. On the horizontal plane the cells are 50 x 50 m, which is suitable for resolving the topographic details. To reduce the number of computational cells while maintaining a suitable model accuracy to capture water table fluctuations, non-uniform size is adopted in the vertical direction: the top two cell layers are each a quarter of the model thickness, and the deepest layer is half the model thickness at the location of the model cell. Each cell layer consists of 686 rows and 727 columns, resulting in about $6.01 \times 10^5$ active grid cells. The model domain extends from north of the town of Abbiategrasso to south of the city of Pavia where the Ticino river joins the Po river, covering an area of 501.5 km$^2$. The surface topography of the model was generated using a 5 m DTM of the region (Regione Lombardia, 2015) resampled to 50 m resolution using the bilinear resampling method in QGIS. Along the river channel, the grid cell surface elevations were increased by 3 m to better represent the elevation of the riverbanks, rather than the elevation of the river at the time the DTM data were measured. The surface elevations of the lateral plains along the edges of the model domain are about 20-30 m higher than the central valley through which the Ticino river flows, resulting in a steep slope between the two zones (Fig. 2b).

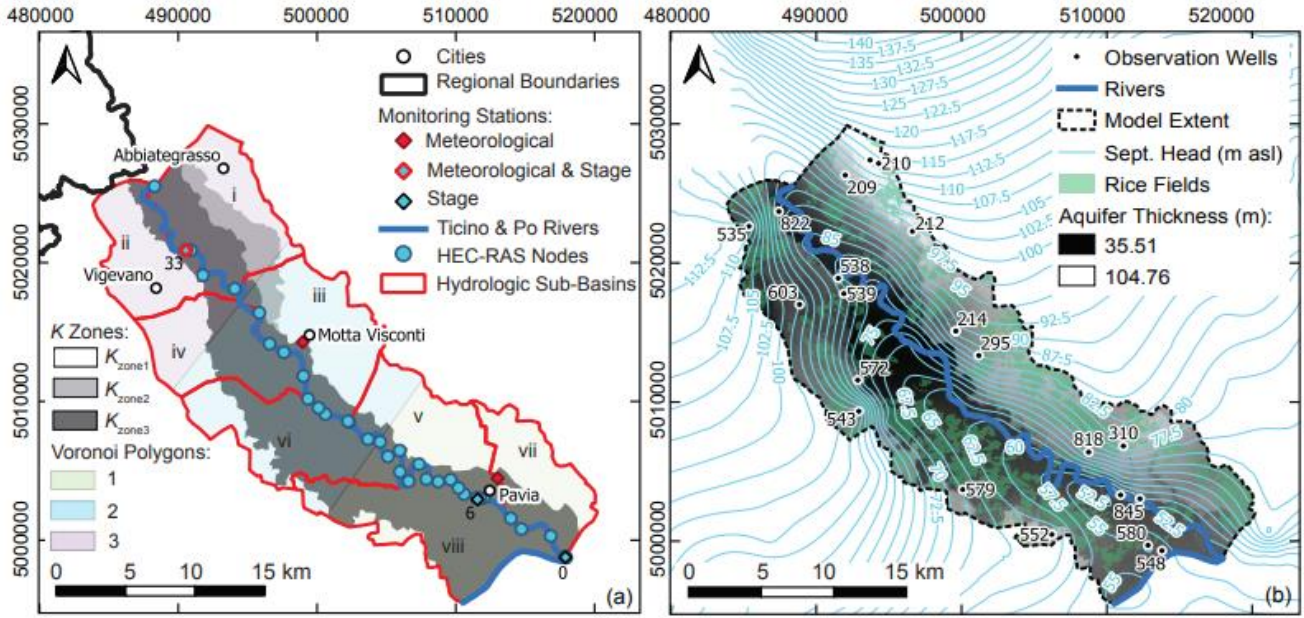

**Figure 2**. Maps depicting information used to create the groundwater flow model and its inputs. (a) Hydraulic conductivity zones, hydrometric and meteorological stations, extent of Voronoi polygons and hydrologic sub-basins (i-viii), and locations of the nodes from the HEC-RAS model used to interpolate the stage values along the rest of the river. (b) The interpolated groundwater level contours of the surficial unconfined aquifer during September 2014, the locations of groundwater wells, the locations of rice fields, and the unconfined aquifer thickness within the study area.

Within the study area, the hydraulic conductivity of the modeled surficial unconfined aquifer tends to be higher in the river valley and lower in the higher elevation plains, with an area with intermediate hydraulic conductivity in the northern portion of the study area between the plain and river valley on the eastern side of the river, resulting in three hydraulic conductivity zones (Fig. 2a). Zone 1 is located on the higher elevation plains, at the eastern and western margins of the model domain. It consists of a mix of sand and gravel (10-20 mm size fraction), along with loam in some locations. The hydraulic conductivity $K_{zone1}$ is estimated to range from $1\times10^{-3}$-$1\times10^{-4}$ m/s in the north to $1\times10^{-4}$-$1\times10^{-5}$ m/s in the south, with an approximate ratio of horizontal to vertical hydraulic conductivity of 5:1. The bottom of this zone is bordered by a clay layer, which serves as an aquitard. Zone 2 has similar hydraulic properties but may have a slightly higher hydraulic conductivity $K_{zone2}$. The upper limit of Zone 2 coincides with the topographic surface, while the lower limit is not accurately known, but is estimated at about 80 m asl near the city of Vigevano. Zone 3 is mainly composed of gravel and sand, resulting in the highest hydraulic conductivity $K_{zone3}$ ranging from $1\times10^{-2}$ to $1\times10^{-4}$ m/s. The upper limit coincides with the topographic surface while the lower limit is around 60-70 m asl around Vigevano. The boundary of Zone 3 with Zone 1 and Zone 2 is almost vertical. These three zones with hydraulic conductivities $K_{zone1}$, $K_{zone2}$, and $K_{zone3}$ were delineated in QGIS both at the surface and at depth. In the top model layer, all three zones are present, according to the description provided. In the second layer, only zones 1 and 3 are present since zone 2 terminates at about 80 m asl near Vigevano as discussed above. The east-west extent of zone 3 in layer 2 was also decreased by 300 m on each side of the river valley to decrease the width of zone 3 at depth. In the third and deepest layer, only zone 1 is present since zone 3 extends to a depth of about 60-70 m asl near Vigevano.

**2.2.2 Rivers**

The Ticino river flows south and slightly east through the center of the study area and converges with the Po river along the southern border of the study area south of the city of Pavia. The length of the Ticino river in the study area is 55.7 km while the length of the Po river along the southern border of the study area is about 9.1 km. The reach of the Ticino river within the model domain is predominantly a gaining river (De Caro et al., 2020). Two gauging stations are located along the Ticino river within the study area near the cities of Vigevano (ARPA gauge SS494) and Pavia (ARPA gauge SS35), while a third station is located at the confluence of the Po and Ticino rivers (AIPo, 2020; Fig. 2a). The stage data from the two gauging stations along the Ticino were acquired from the Agenzia Regionale per la Protezione dell'Ambiente (ARPA) online data portal (ARPA Lombardia, 2020a), while the stage data from the confluence were acquired from the Agenzia interregionale per il fiume Po monitoring network at the Ponte della Becca gauge (AIPo, 2020). Throughout most of the year, stage values fluctuate within a narrow range of only a few meters, except during November. During September 2014, the stage values fluctuate by less than a meter over the whole month at all three gauging stations, with the largest range in September stage values of 0.84 m occurring at the confluence of the Po and Ticino rivers and the smallest range of 0.22 m at the Vigevano gauging station (Fig. 3a). It is difficult to compare the stream flow rates at the Vigevano and Pavia gauges due to the limitations of the rating curves and the numerous irrigation withdrawals and artificial inflows along the 37.5 km reach between the two locations.

Both rivers are defined as specified head boundaries (CHD package) in MODFLOW. The head values along the Ticino river were calculated using a 1D unsteady hydraulic model of the river developed with the open-source software HEC-RAS 5.0.7 (HEC-RAS, 2019). The HEC-RAS model extends from the A4 highway bridge, near the town of Magenta, to the confluence with the Po river and is composed of 42 surveyed cross sections (AIPo, 2004; AIPo, 2005). The model was simplified by considering only two Manning roughness coefficients, one for the main channel (0.028) and one for the floodplain (0.1) where the coefficient for the floodplain is always higher, a common choice for such models (Pappenberger et al. 2008, Saleh et al. 2013). During low flow periods, like those during August/September 2014, the upstream boundary condition is a flow hydrograph calculated from the stage levels recorded at Vigevano SS494 gauging station and converted into flow rates using the rating curve proposed by ARPA (ARPA Lombardia, 2020b). During floods the ARPA rating curve is no longer valid and so it is necessary to use the Lake Maggiore discharge flow appropriately shifted over time. Additionally, the Po river exerts a great influence on the water levels of the Ticino river, and so it is necessary to use different downstream boundary conditions depending on the Po stage. During floods, a stage hydrograph is set based on Ponte della Becca gauge station measurements, while Normal Depth is chosen during low flow periods when the backwater effect is negligible. The Vigevano SS494 and Pavia SS35 ARPA gauging stations were used for HEC-RAS model calibration and to confirm the accuracy of the calculated stage values. The HEC-RAS model tends to slightly overestimate the stage during higher flow periods, with maximum stage errors of 0.27 m at the Pavia station and 0.26 m at the Vigevano station during September 2014. The root mean squared errors (RMSE) of the calculated stage values at the two stations are small (Pavia: 0.12 m; Vigevano: 0.14 m). During September 2014, calculated stage values at any HEC-RAS stream node along the Ticino river varied by up to 0.83 m. The calculated stage values at 35 of the HEC-RAS nodes were averaged to obtain an average monthly stage at each of the nodes for September 2014. These average stage values were then used to estimate the average September stage values in each MODFLOW river cell through linear interpolation between each set of adjacent HEC-RAS nodes (Fig. 3b). The average distance between the HEC-RAS computational nodes was 1.56 km with a standard deviation of 0.89 km.

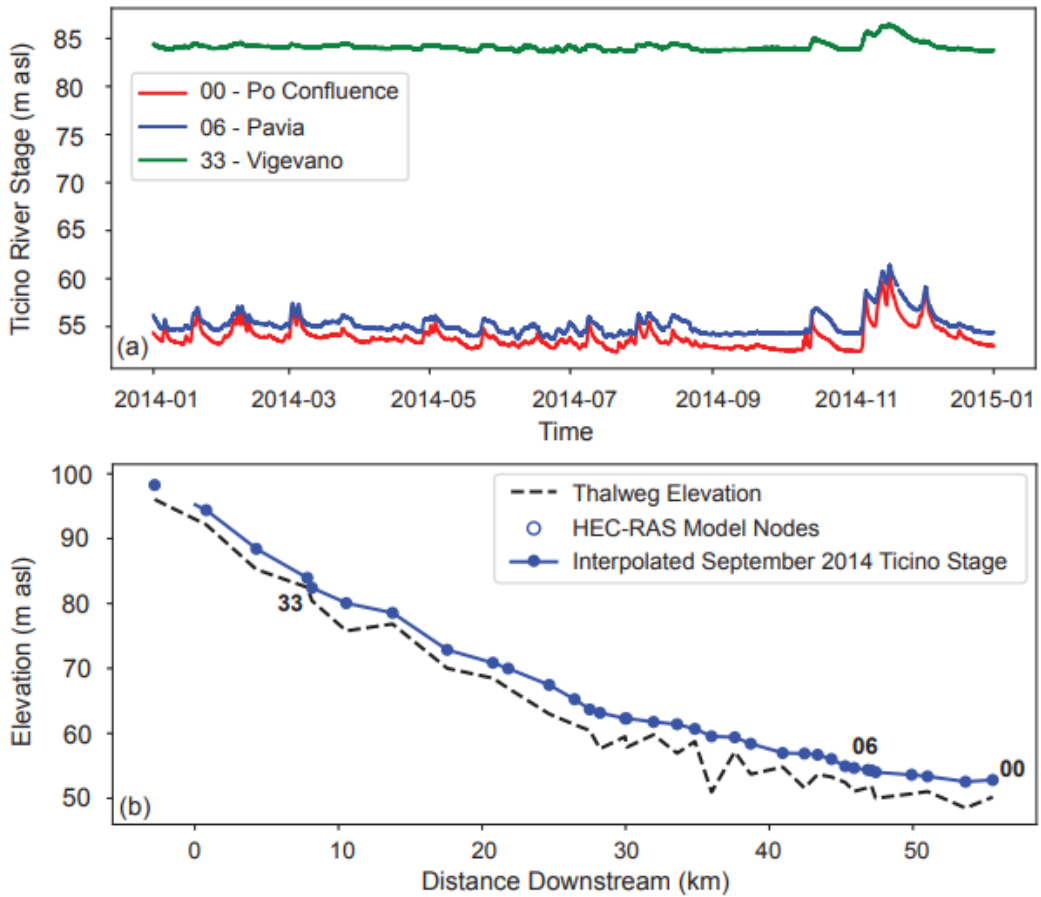

**Figure 3.** (a) Observed stage values at the gauging stations along the Ticino river during 2014. (b) Average September 2014 stage values interpolated along the Ticino river, along with thalweg elevations. The blue dots depict the locations of the HEC-RAS computational model nodes used to obtain the stage values along the Ticino river.

The only gauging station in the MODFLOW model domain along the Po river is located at the confluence of the Po and Ticino rivers. The assumption was made that the stage is constant along the reach of the Po river (9.1 km) at the southern edge of the model domain. Therefore, the stage values at the confluence of the Ticino and Po rivers were used for the whole reach of the Po river in the model domain. While this assumption could introduce some head errors in grid cells adjacent to this river, the errors are likely small and do not affect results further from the boundary. These errors are likely small since the water levels in the most downstream portion of the Ticino river are almost constant between Pavia and the confluence (Fig. 3b Ticino River stage between nodes 00 and 06) and so it can reasonably be assumed that the same occurs in the Po river.

**2.2.3 Groundwater Levels**

A contour maps of the piezometric surface of the unconfined aquifer during September 2014 is available online through the Geoportale della Lombardia (Fig. 2b; Regione Lombardia, 2014; Éupolis Lombardia, 2015). The contour map was created by interpolating the groundwater heads measured at 21 wells during late August/September of 2014 (contour lines in Fig. 2b). The groundwater wells used to create the 'September' 2014 contour map were sampled between August 27 – September 18, 2014, when steady boundary conditions can be reasonably assumed. The groundwater head data used to create the contour map were provided by Regione Lombardia and additional head measurements were provided by ARPA, resulting in measured head data at a total of 22 monitoring wells within the study area (Fig. 2b). The groundwater flow model is calibrated to this August/September 2014 data at the 22 monitoring wells.

The September 2014 contour map of groundwater heads is used to set the head values at the model edges. The western and eastern edges of the model domain approximately coincide with the edges of the Ticino groundwater basin to align with the groundwater divide (Fig. 1; ARPA Lombardia, 2020b) and help minimize error introduced by the uncertainty at the model edges. The northern model edge has been chosen to represent a flux line intersecting the groundwater head contour at a right angle, such that groundwater flow is parallel to the northern edge of the model domain. In this way, there is no need to define groundwater flow rate through the northern edge of the domain. A portion of the north-eastern edge follows the 117.5 m contour line to transition from the northern to eastern edge (Fig. 2b). All model edges that are not aligned with the rivers are parameterized as general head boundaries (GHD) using the groundwater levels from the September 2014 piezometric contour map (Regione Lombardia, 2014).

**2.2.4 Springs & Fontanili**

About 140 groundwater springs are mapped within the study area (Regione Lombardia, 2007a; Regione Lombardia, 2013b; Magri, 2020; Gardini, 2021). Springs typically occur along the transition zone from the higher plains to the river valley where the steep surface topography intersects the unconfined aquifer and the hydraulic conductivity of the aquifer increases (Fig. 1b, 2a; Balestrini et al., 2021). Historically, springs were often enlarged and fitted with pipes to enhance the flow of groundwater to the surface for human use (De Luca et al., 2014; Balestrini et al., 2021). Such modified springs are called fontanili. Water flowing out of springs is often routed through irrigation ditches and the canal network for irrigation purposes. Previous studies have measured flux rates from fontanili ranging from less than 4 to 80 l/s (0.004-0.08 $m^3$/s; De Luca et al., 2014) and from 10 to 1000 l/s (0.01-1.0 $m^3$/s) or more (Fumagalli et al., 2017). In a study with data from 2016 and 2017, fontanili discharges ranged from 1.1 – 274 l/s (median of about 30 l/s), with the largest mean discharges during July and August followed by a decrease in discharge during the autumn and winter (Balestrini et al., 2021).

Springs and fontanili are incorporated as drain boundaries within the model, similar to the approach taken by other studies (e.g., Bianchi Janetti et al., 2019). The approximately 140 springs and fontanili occur within 132 model grid cells, to which drain boundaries are assigned. The drain boundaries remove groundwater from the model when it reaches the elevation of the land surface. The rate of discharge from the grid cells assigned drain boundaries is proportional to the difference between the hydraulic head and land surface elevation in that grid cell. The proportionality constant used to determine the discharge rate from the grid cells containing drain boundaries is termed the drain conductance. In a study in a nearby basin, assigned drain conductance values ranged from 0.03 - 85.1 $m^2$/s (Musacchio et al., 2021). The effect of the uncertainty in the value of the drain conductance is explored using the Morris Method in subsequent analyses. The modeled discharge rates from the drain boundaries are assessed to ensure that the flux is reasonable compared to measured discharges from springs and fontanili in the region, and the percentage of drain boundary cells with active discharges is calculated to determine how many of the known spring locations behave as springs within the model.

**2.2.5 Aquifer Recharge**

During the growing season, recharge to the surficial aquifer comes from precipitation, infiltration of irrigation water from unlined canals (Pognant et al., 2013; Clemente et al., 2015) and flood irrigation within rice fields (Rotiroti et al., 2019; Lasagna et al., 2020). Previous studies in the Po Plain demonstrate that recharge due to precipitation is insufficient to explain the observed pattern in groundwater levels (Rotiroti et al., 2019; Lasagna et al., 2020) and that 55-88% of aquifer recharge is due to irrigation (Rotiroti et al., 2019). Aquifer recharge from both precipitation and irrigation activities is applied to the current groundwater flow model using the MODFLOW recharge (RCH) boundary. The locations of the canals and irrigation ditches were obtained from a shapefile downloaded from Geoportale della Lombardia (Regione Lombardia, 2007a) while the locations of rice fields (Fig. 2b) were obtained from the DUSAF 6.0 land use map (Regione Lombardia, 2019). The recharge rate due to precipitation was calculated by subtracting the estimated runoff and evapotranspiration rates from the precipitation rate in each hydrologic sub-basin. Meanwhile, the recharge rate due to irrigation activities was estimated through model calibration. Where rice fields or irrigation canals are present, the underlying model grid cells receive recharge both due to precipitation and irrigation. While the ponded irrigation waters in rice fields are often perched above the water table, sometimes the piezometric surface can reach the land surface depending on site specific conditions, connecting the aquifer system to the flooded rice fields. To simulate this behavior, the drain boundary condition is also implemented in cells containing rice fields to allow water levels in rice fields to reach levels slightly above the land surface since a water level of 5-10 cm is maintained in the rice fields during flood irrigation. Recharge that exceeds this ponding depth is removed from the model, as excess recharge would similarly be removed as runoff from the rice fields back into the irrigation canals.

Meteorologic, land use, and soil data are used to calculate the evapotranspiration (ET) and runoff rates used to estimate the recharge rate due to precipitation. To account for spatial variation of the hydrologic and meteorologic characteristics, the study area was divided into eight sub-basins (Fig. 2a) based on the watersheds of the minor tributaries and the morphology of the territory. Runoff was calculated using the Soil Conservation Service (SCS) Curve Number (CN) method (Mishra & Singh, 2003) to obtain a maximum potential infiltration rate. Then, the potential evapotranspiration (PET) rate was calculated using the FAO-56 Penman-Monteith method (Allen et al., 1998). Therefore, recharge due to precipitation ($R_P$) is calculated as:

$$R_P(t) = P(t) - P_e(t) - PET(t) \qquad [2]$$

where:
- $P(t)$ = Precipitation: cumulative rainfall height at time $t$ [mm];
- $P_e(t)$ = Runoff: cumulative precipitation excess at time $t$ [mm];
- $PET(t)$ = Penman-Monteith potential evapotranspiration rate at time $t$ [mm].

If $PET(t)$ exceeds $P(t) - P_e(t)$ then $R_P$ equals zero and no recharge occurs during that period. The precipitation, runoff, and potential evapotranspiration rates were calculated on a weekly timescale, resulting in estimates of the weekly recharge rate [mm/week]. Rates were estimated over weekly periods to obtain more accurate estimates of ET and runoff. When recharge is estimated on a daily timescale the initial abstraction within the runoff calculation is performed too frequently, ignoring the pre-existing soil moisture content. When recharge rates are calculated over a monthly period, too much precipitation leaves the system as evapotranspiration since this method assumes precipitation that occurred anytime throughout the month is available to be lost through ET even though the high permeability of the soil allows for precipitation to infiltrate the subsurface over a shorter timescale. The weekly estimated rates result in a more realistic frequency of initial abstractions and timeframe over which PET can act on the precipitation. Therefore, the weekly runoff and PET rates were used in the current study to estimate the amount of recharge that occurred during the study period. The weekly recharge rates were then summed to obtain the monthly recharge rate [mm/month]. The August and September 2014 recharge rates were then averaged to obtain a recharge rate for the study period, since the observation wells were measured during a period spanning both months and infiltration usually takes less than a month to influence groundwater levels (Lasagna et al., 2020).

The runoff for each sub-basin as a function of cumulative precipitation, soil characteristics and land use was estimated using the Soil Conservation Service (SCS) Curve Number (CN) method (Mishra & Singh, 2003). According to this method, the excess precipitation (runoff) is calculated as:

$$P_e(t) = \frac{(P(t)-I_a)^2}{(P(t)-I_a+S_\infty)} \qquad [3]$$

where:

    $P_e(t)$ = cumulative precipitation excess (runoff) at time $t$ [mm];
    $P(t)$ = cumulative rainfall height at time $t$ [mm];
    $I_a$ = initial abstraction (initial loss) [mm];
    $S_\infty$ = potential maximum retention, a measure of the ability of a watershed to abstract and retain storm precipitation [mm].

Eq. 3 is valid where the cumulative rainfall height exceeds the initial abstraction, i.e. $P(t) > I_a$. Assuming $I_a = 0.2\ S_\infty$, the only parameter is $S_\infty$ which can be estimated using the equation:

$$S_\infty = 25.4 \cdot [1000/CN - 10] \qquad [4]$$

In Eq. 4, CN is an empirical scalar parameter that incorporates all the hydrological characteristics of the soil and determines its tendency in generating runoff. The soil is divided into four hydrological groups (A, B, C, D) based on the capability of the terrain to absorb the rainfall, decreasing from Group A (low runoff potential) to Group D (high runoff potential). The soil group subdivision for the study area was carried out based on the texture and the granulometry of the superficial soil layer (USDA Textural Soil Classification) available in the Pedological Map (1:250000) of Lombardy (Regione Lombardia, 2013a). Most of the area (83%) is characterized by soils with high/moderate infiltration rates corresponding to Group A/B while a minor portion (17%) has a lower infiltration rate (Group C). Soil group A (19%) is characterized by well-drained sand and a gravel skeleton and is concentrated in the central zone along the river, while soil Group C (silt loam) is mainly distributed along the northern boundary of the study area and close to the southern confluence between the Ticino and Po rivers. Each land cover type corresponds to a CN value depending on the hydrological soil class, defined in specific tables widely available in the literature. For this study, a conversion table proposed by Castelli (2014) is used to transform the Corine Land Cover codes available in the DUSAF 6.0 land use map (Regione Lombardia, 2019) into CN values. Then the final CN value for each sub-basin was calculated with an area-weighted average based on the hydrological soil classes cover (Table 1). The curve numbers were used in the SCS Runoff Curve Number method to estimate the amount of runoff generated over week-long periods in each of the 8 sub-basins.

Potential evapotranspiration (PET) was calculated using the FAO-56 Penman-Monteith method which considers the air temperature, wind speed, solar radiation, and relative humidity (Allen et al., 1998). This meteorological data was accessed through the online data portal managed by ARPA, which maintains a network of meteorological stations in the Lombardy region (ARPA Lombardia, 2020a). Three of the meteorological stations are located within the study area and recorded meteorological data during 2014 (Fig. 2a). Precipitation, relative humidity, shortwave radiation, wind speed, and air temperature were measured every ten minutes (Fig. 4). However, not all data types were available from all stations since some stations were not equipped with all sensors. The Motta Visconti station, which was the most centrally located in the study area, recorded all five data types during 2014. The Pavia Folperti station was missing wind speed data. The Vigevano station lacked wind speed, relative humidity, and solar radiation data. Daily average air temperatures range from below -1°C to above 28°C during 2014 (Fig. 4a), with summer daily average temperatures between about 18 to 28°C. The daily relative humidity in the study area is typically greater than 60% (Fig. 4b), the daily solar radiation is highest during the summer (Fig. 4c), and the daily wind speed is typically between 0-2 m/s (Fig. 4d). The study area received 1310 mm of precipitation during 2014. September had the least precipitation with 10.0 mm. The average weekly precipitation rate during August 2014 was 25.6 mm/week (3.7 mm/day) while the average weekly precipitation rate during September 2014 was 2.2 mm/week (0.3 mm/day) (Fig. 4e, 4f). The precipitation data also show that the Pavia Folperti station typically receives the least precipitation, while the Vigevano station receives the most, indicating precipitation increases in the basin from south to north. Precipitation data from 2002 to 2019 indicate most precipitation typically occurs during November. More precipitation occurred in 2014 than the annual average, with the Motta Visconti, Pavia, and Vigevano meteorological stations measuring 1263.2, 1291.8, and 1390.2 mm ($\sigma$ = 269.0, 273.3, 508.9 mm) of precipitation during this year as compared to their annual averages of 822.4, 746.1, and 964.9 mm (n = 10, 7, 12 years of data for the three stations). This represents over a 50% increase in precipitation on average at the three stations during 2014 relative to the average annual precipitation. During 2014, many months (e.g., January, February, June, July, August, and November) received substantially more precipitation than average, while May and September received substantially less precipitation than average.

      Weekly PET rates were calculated at each of the three meteorological stations (Fig. 2) using the FAO-56 Penman-Monteith method implemented in the PyETo package in Python (Richards, 2015). Then Voronoi polygons were used to calculate the spatially weighted PET and precipitation rates within each sub-basin in the study area (Fig. 2a). The Motta Visconti meteorological station is centrally located in the study area, and descriptive statistics showed that the weather data at the other two meteorological stations (Pavia Folperti & Vigevano) are more similar to the weather data at the Motta Visconti station than to each other. Therefore, weather data from the Motta Visconti station were substituted for data that were missing at the other two meteorological stations to estimate PET. August/September 2014 had an average weekly PET rate of 22.5 mm/week (3.2 mm/day), ranging from 21.8 to 23.4 mm/week depending on the sub-basin.

      Using the methods described above, the estimated average recharge rate is 10.9-33.4% of the weekly precipitation rate across the sub-basins, with an area weighted average over the whole model domain of 21.6% of the average precipitation rate during August/September 2014. Average weekly recharge rates vary from 1.7 to 5.9 mm/week depending on the sub-basin, with an average across the sub-basins of 3.5mm/week (0.49 mm/day) during the study period. The remainder of the precipitation is either removed through evapotranspiration or surface runoff, with runoff representing only a small portion of the budget due to the flat nature of most of the land surface within the study area. Average weekly August/September 2014 estimated runoff rates vary from 4.7-22.9% of the precipitation rate depending on the sub-basin, with an average runoff of 10.3% of the precipitation. Meanwhile, the average weekly estimated evapotranspiration rate is 68.1% of the precipitation during the study period. Therefore, most of the precipitation during August/September 2014 leaves the system as evapotranspiration, while only a small portion of the precipitation leaves the system as runoff or is available to recharge the surficial aquifer (Fig. 4f).

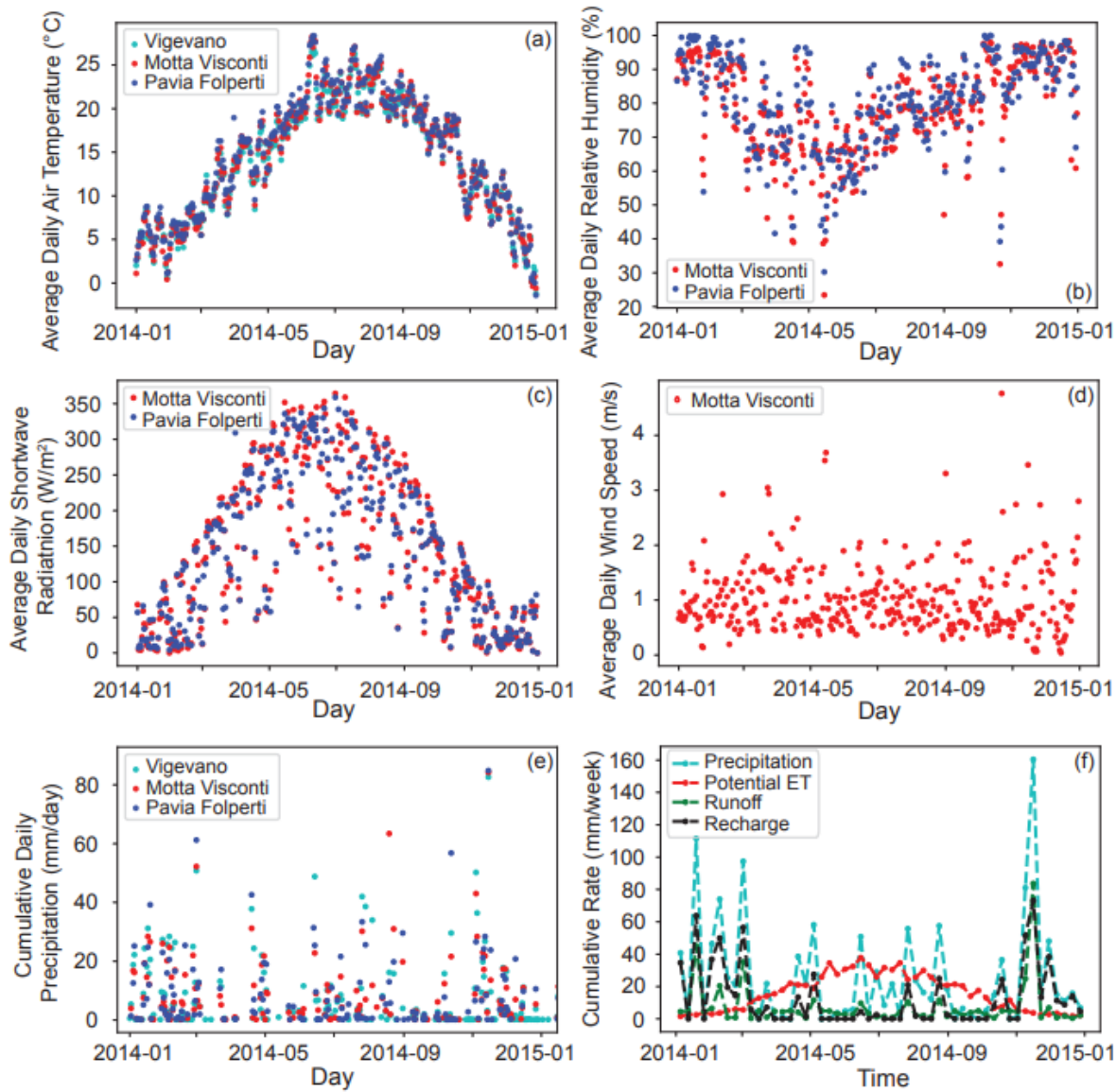

**Figure 4**. (a-e) Meteorological data recorded during 2014 at the Vigevano (cyan), Motta Visconti (red), and Pavia Folperti (blue) monitoring stations. The (a) air temperature (°C), (b) relative humidity (%), (c) shortwave radiation (W/m$^2$), and (d) wind speed (m/s) data are averaged at a daily timescale. (e) The daily cumulative precipitation rate (mm/day) at each station. (f) The weekly cumulative precipitation, PET, runoff, and recharge due to precipitation averaged over the study area according to the hydrologic sub-basins and Voronoi polygons.

### 2.3 Morris Method Sensitivity Analysis

A sensitivity analysis was performed using the Morris Method (Morris, 1991) to assess the sensitivity of the groundwater model to the uncertainty in the model input parameters. The Morris Method calculates the effect of changing the value of a single model input parameter on the chosen model output. Roughly speaking, the Morris Method consists in evaluating the (finite difference approximation of the) gradient of the model outputs with respect to model parameters at several random locations in the parameters space, and then computing the average of such gradients to determine which parameters are overall more influential. More precisely, the value of each model parameter ($p_i$) is adjusted one-at-a-time along a single trajectory ($j$) in the parameter space of the model until the values of each parameter have been changed. Then a new starting point is chosen from which each parameter is again adjusted, creating another parameter trajectory. After running the model using the parameter combinations generated by each of numerous trajectories, the effects of each parameter value on the chosen model output (e.g., modeled head, groundwater flux), termed the elementary effects (*EE*), can be calculated as:

$$EE_{p_i}(j) = \frac{f(p_1,...,p,+\Delta,...,p_N) - f(p)}{\Delta} \qquad [5]$$

These effects are then averaged to calculate the mean elementary effect (*μ*) of a parameter (Morris, 1991; Campolongo et al. 2007; Ruano et al., 2012; Bianchi Janetti et al., 2019). The larger the value of *μ* for the considered parameter, the more sensitive the model output is to the value of that parameter. The absolute value of the mean elementary effect (*μ**) is also calculated to account for effects with opposite signs (such as non-monotonic behavior) which would lower the value of the elementary effect compared to the mean elementary effect, decreasing the perceived influence of the parameter (Campolongo et al. 2007). The absolute value of the mean elementary effect (*μ**) is calculated as:

$$\mu^*_{p_i} = \frac{1}{r}\sum_{j=1}^{r}|EE_{p_i}(j)| \qquad [6]$$

where *r* is the number of trajectories. The mean elementary effects for all the parameters can then be compared to rank the relative influence of the parameters (Campolongo et al. 2007; Reinecke et al., 2019). The standard deviation of the elementary effects (*σ*) is also calculated to determine whether the effect of the parameter value on the model output is linear, or whether interactions occur with other parameters: if the *σ* is large relative to *μ**, the relationship between a change in the parameter value and the resulting model output is either nonlinear and/or the parameter interacts with other model parameters and so the elementary effect is affected by the values of the other input parameters (Morris, 1991; Feng et al., 2019; Reinecke et al., 2019).

The SALib package (Herman & Usher, 2017) in Python was used to conduct the Morris analysis to determine the relative sensitivity to each uncertain input parameter of: (*i*) the overall RMSE of the heads at the observation wells (RMSE$_h$); (*ii*) the percentage of model cells with calculated heads above the land surface (H$_{PAS}$); and (*iii*) the groundwater flux rate to the stream (GW Flux). Additionally, the sensitivity of the calculated heads at each of the observation wells to the model input parameters was analyzed. The input parameters considered for this analysis were the hydraulic conductivity of each of the three zones ($K_{zone1}$, $K_{zone2}$, $K_{zone3}$), the aquifer recharge rate due to rice field irrigation ($R_{Irrig}$; recharge due to leakage from irrigation canals is also included in this term), the river stage ($S_{Riv}$), the groundwater head at the general head boundaries ($H_{GHB}$), and the conductance of the drain cells ($C_D$) (Table 2). These 7 uncertain parameters were each sampled at 6 equispaced values across their range of uncertainty, resulting in a total of about $6^7$ possible parameter trajectories. For the parameters with uncertain ranges spanning orders of magnitude ($K_{zone1}$, $K_{zone2}$, $K_{zone3}$, $R_{Irrig}$, $C_D$), the exponents of the parameters were uniformly sampled in log space to appropriately represent the whole parameter space (Reinecke et al., 2019). The range of uncertainty for each parameter value and the sampled values are shown in Table 2. 1000 parameter trajectories were then randomly generated from the possible trajectories and sets of trajectories (n = 30, 50, 70, 100 and 150) were then chosen from these generated trajectories by selecting those at a maximum distance from one another within the parameter space using the methods of Campolongo et al. (2007) and Ruano et al. (2012) in SALib. Morris analyses were conducted with different numbers of sets of trajectories to assess the stability of the ranking of the parameter significance since the number of sets of trajectories can influence the ranking and the optimal number of trajectory sets differs due to the model (Ruano et al., 2012). Parameter trajectories for which a model run did not converge to a stable solution due to non-physical parameter value combinations were removed since the trajectories are independent (Campolongo et al., 2007). The outputs of the Morris method calculated in SALib include the mean elementary effect ($\mu$), the absolute value of the mean elementary effect ($\mu^*$), and the standard deviation of the elementary effect ($\sigma$) (Campolongo et al. 2007; Ruano et al., 2012). These values of $\mu^*$ are then used to determine which parameters are most influential. Meanwhile, parameters with minimal influence on the model results can be set to a constant value.

**2.4 Parameter Optimization**

Brute-force optimization (i.e., an exhaustive search over a large, predetermined set of candidate points over the entire parameter space) was subsequently performed, followed by an actual optimization (the 'optimize.minimize' function in Python using the Nelder-Mead method) using the brute-force results as a starting point, to determine an optimal set of model input parameters through the minimization of a joint error metric that will be detailed in the following. The initial brute-force optimization step was introduced since the plain use of standard minimization procedures resulted in the convergence to numerous local minima, rather than a clear global minimum. The set of candidate points for the brute-force step was constructed as a Cartesian grid over the parameter space, choosing a set of values for each input parameter and then taking every combination of such values. Based on the results from the Morris sensitivity analysis (see Section 3.2 later on), the multidimensional grid consisted of 15 values of $K_{zone1}$ and $K_{zone3}$, 4 hydraulic conductivity values of $K_{zone2}$, and 20 recharge rates, $R_{Irrig}$, due to irrigation activities (Table 3). Parameter combinations that did not satisfy the constraints on the system ($K_{zone3}$ is always greater than or equal to that of $K_{zone1}$ and $K_{zone2}$, and $K_{zone2}$ is always greater than or equal to that of $K_{zone1}$) were excluded from the analysis. Meanwhile, the values of $S_{RIV}$, $H_{GHB}$, and $C_D$ were set constant since the uncertainty in these parameters were found to have a minimal influence on the model results. $S_{RIV}$ and $H_{GHB}$ were set to their measured/interpolated values, while $C_D$ was set to 100 m$^2$/s. These constraints resulted in a multidimensional grid of 6300 input parameter combinations for which model outputs were calculated. In the early phase of this work, the optimal set of model input parameters was selected by determining which set of input parameter values minimized the RMSE of the heads at the observation wells. However, running the model for such calibrated values of the parameters resulted in unrealistic predictions (i.e., the percentage of cells with calculated heads above the land surface was too large). Therefore, a joint error metric was instead minimized to determine a set of model input parameters that produce modeled heads that match the available data while being physically consistent. We briefly sketch here the main steps of the procedure and refer the reader interested in the mathematical details to Piazzola et al. (2021) and Porta et al. (2014), which also demonstrate additional applications of the generalizable method.

Following Carrera & Neuman (1986) and Nocedal & Wright (1999), the joint error metric considered was based on a classical Bayesian inversion approach, where RMSE minimization is replaced by minimization of the so-called Negative Log Likelihood (NLL) whose role is to measure the plausibility of the experimental measures given values of the model parameters. The NLL is expressed as:

$$NLL = \frac{1}{2\sigma_h^2}\sum_{i=1}^{n_{wells}}(h_i - h_i^*)^2 + \frac{1}{2\sigma_{H_{PAS}}^2}(H_{PAS} - H_{PAS}^*)^2 + n_{wells}\, log(\sigma_h) + log(\sigma_{H_{PAS}}) + \left(\frac{n_{wells}+1}{2}\right)log2\pi \qquad [7]$$

Here, $h_i$ and $h_i^*$ are the modeled and measured heads at each of the observation wells ($n_{wells}$ = 24) respectively, $\sigma_h$ is the standard deviation of the measurement error in the heads of the wells (also unknown and to be determined within the optimization procedure), and $H_{PAS}$ and $H_{PAS}^*$ are the modeled and nominal/reference percentages of model grid cells with calculated heads above the land surface respectively. Finally, $\sigma_{H_{PAS}}$ is a value fixed a-priori that identifies a range of plausible values for the modeled $H_{PAS}$ given the nominal value $H_{PAS}^*$. More specifically, the plausibility of $H_{PAS}$ is assumed to behave as a gaussian function centered at $H_{PAS}^*$, with a standard deviation equal to $\sigma_{H_{PAS}}$. Setting $H_{PAS}^*$ to 1.0% and $\sigma_{H_{PAS}}$ to 0.3% results in the plausibility of $H_{PAS}$ less than 0 being outside the interval 0%-2.0% (yielding a reasonable number of active springs and fontanili). Note that the coefficients $\frac{1}{2}\sigma_h^2$ and $\frac{1}{2}\sigma_{H_{PAS}}^2$ can be also interpreted as weights that establish the relative importance of the two terms (RMSE of the well heads and deviation from the expected $H_{PAS}$) in the minimization procedure (i.e., whether the optimization should prioritize the well measurements or the percentage of flooded land surface). The Bayesian approach thus can be seen as advantageous because it provides us with a systematic way to adjust these coefficients to our a-priori knowledge: in particular, $\sigma_{H_{PAS}}$ is tuned such that it is highly unlikely that the method will deliver optimized parameters that would predict a flooded area outside of the interval 0%-2.0% of the total surface. Model simulations where more than approximately 2% of grid cells have calculated heads above the land surface are considered unrealistic since groundwater heads should only be above the land surface in cells where springs and fontanili are located. The brute-force optimization with respect to the model input parameters was repeated for 50 equispaced values of $\sigma_h$, ranging from 0.98 to 4.5, and the 15 combinations of parameters and $\sigma_h$ that yielded the overall smallest value of Eq. 7 were selected. Each of these 15 combinations were then used as the starting inputs for the actual optimization algorithm to further refine the estimated parameter values.

      Estimates of the robustness/reliability of the optimized parameter values were then calculated by assuming that the values of the parameters after optimization still have some residual uncertainty and follow a joint Gaussian distribution, centered at the optimized values and with a covariance matrix equal to the inverse of the Hessian of the NLL at the optimized value (in particular, we recall that the square root of the diagonal entries of such a covariance matrix represents the standard deviations of the residual uncertainties of the parameters). This construction of the covariance matrix quantifies the intuitive fact that if the NLL has a narrow minimum, there is little uncertainty left on the values of the parameters, whereas conversely a shallow NLL implies that its value would not change considerably even by modifying substantially the values of the parameters, and therefore there is a large uncertainty on the values of the parameters even after optimization. Ad hoc simplified formulas for the Hessian matrix are available in the literature and have been employed in this work, see again Piazzola et al. (2021) and Porta et al. (2014) for details. Lastly, to confirm the reliability of the resulting parameter set selected by the NLL optimization procedure, the percentage of active drain boundaries in grid cells containing mapped springs/fontanili, the magnitude of the fluxes out of the active drains, and the net groundwater flow into the Ticino river (the river is known to be a predominantly a gaining river; De Caro et al., 2020) were all checked to confirm that their values were reasonable.

## 3. Results
### 3.1 Aquifer Recharge due to Precipitation

      The aquifer recharge rate due to precipitation was directly quantified using the SCS Runoff Curve Number method to estimate runoff and the Penman-Monteith method to calculate the potential evapotranspiration on a weekly basis and then averaged over the August-September 2014 study period. The estimated recharge rate due to precipitation ranged from $2.78 \times 10^{-9}$ to $9.73 \times 10^{-9}$ m/s depending on the hydrologic sub-basin. Mean precipitation rates during the period of study (August & September 2014) ranged from $2.23 \times 10^{-8}$ to $2.98 \times 10^{-8}$ m/s depending on the hydrologic sub-basin. Therefore about 10.9% to 33.4% (mean of 21.6%) of the precipitation within each sub-basin becomes aquifer recharge, with the hydrologic sub-basins containing the cities of Abbiategrasso and Vigevano having the lowest recharge rates due to precipitation (14.6% and 10.9% respectively). A study by Canepa (2011) estimated recharge

rates within the Po Plain between the Ticino and Oglio rivers during 2005. Within the area encompassed by the current study, Canepa (2011) estimated annual recharge rates due to precipitation to be about 27.4-46.6% of the annual precipitation rate depending on location, but states that these are overestimates because it is cooler and rainier further north at the weather station used in the study, resulting in greater estimated recharge rates. Recharge is small to nonexistent during the summer months when high air temperatures lead to large evapotranspiration rates. Similar to the estimates in the current study, Canepa (2011) estimates that 25% or less of the precipitation during August and September becomes aquifer recharge. Therefore, the average estimated recharge rate of 21.6% due to precipitation input into the groundwater flow model is realistic.

### 3.2 Morris Method Sensitivity Analysis

The Morris Method was used to analyze the relative sensitivity of three model metrics ($RMSE_h$, $H_{PAS}$, GW Flux) to seven uncertain model parameters ($K_{zone1}$, $K_{zone2}$, $K_{zone3}$, $R_{Irrig}$, $S_{Riv}$, $H_{GHB}$, $C_D$). As explained in Section 2.3, sets of 30, 50, 70, 100 and 150 parameter trajectories were analyzed to ensure that the rankings of the parameters were stable. The parameter ranking is generally stable across all sets of parameters for the three metrics (Fig. 5), with only slight changes in the parameter rank when two parameters have very similar mean elementary effects ($R_{Irrig}$ and $K_{zone3}$, Fig. 5a; $K_{zone1}$, $K_{zone3}$, and $R_{Irrig}$ Fig. 5b). These exceptions do not significantly change the overall behavior of the parameters in terms of relative influence. We also report the ranges of calculated heads obtained from model runs used for the Morris Method (Fig. 5d); for most wells, these ranges include the measured values, and for the few cases where the measured value is outside the range, it is not dramatically different. This observation supports that our model and the selected parameter ranges are able to capture the behavior of the system.

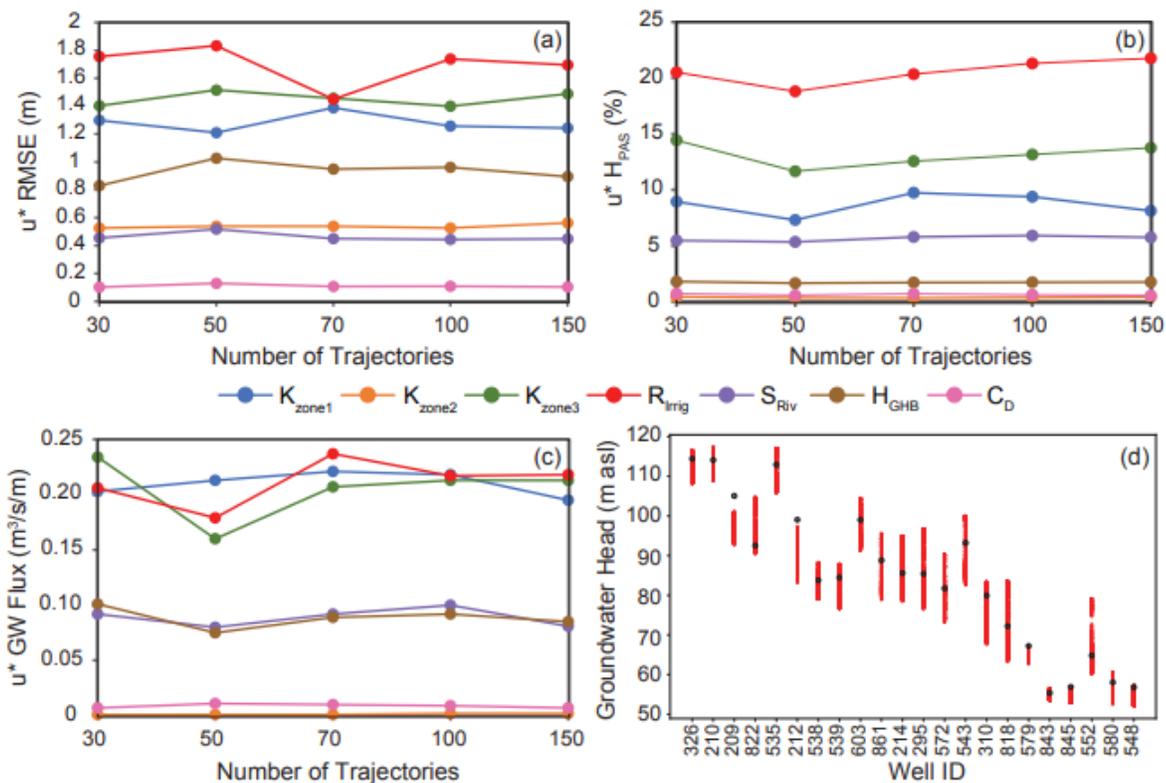

**Figure 5**. The absolute value of the mean elementary effect ($\mu^*$) of each model parameter for an increasing number of parameter trajectories. (a) $\mu^*$ of the RMSE of the heads at the observation wells. (b) $\mu^*$ of the groundwater flux rate to the Ticino river.

(c) $\mu^*$ of the percentage of model grid cells with calculated groundwater heads above the land surface. (d) Range of modeled heads from the Morris sensitivity analysis (red points) compared to the measured head data at the observation wells (black circles). Input parameters include the hydraulic conductivity values of the three aquifer zones ($K_{zone1}$, $K_{zone2}$, $K_{zone3}$), the recharge rate due to irrigation of rice fields ($R_{Irrig}$), the stage of the Ticino River ($S_{Riv}$), the groundwater head at the general head boundaries ($H_{GHB}$), and the conductance of the drain boundaries ($C_D$).

The $RMSE_h$ was most sensitive to the aquifer recharge rate due to irrigation, followed by $K_{zone3}$ and $K_{zone1}$ (Fig. 5a; Fig. 6a). The groundwater flux rate to the stream was most sensitive to the recharge rate due to irrigation, $K_{zone3}$ and $K_{zone1}$ (Fig. 5b; Fig. 6b). The percentage of model cells with calculated heads above the land surface was most sensitive to the recharge rate due to irrigation, followed by $K_{zone3}$ and then $K_{zone1}$ (Fig. 5c; Fig. 6c). All three metrics were insensitive to the uncertainty in the drain conductance, while both the groundwater flux rate and $H_{PAS}$ were insensitive to $K_{zone2}$ (Fig. 5; Fig. 6).

When $RMSE_h$ is the metric, the elementary effects of $K_{zone1}$, $K_{zone3}$ and $R_{Irrig}$ have large standard deviations values ($\sigma$ = 1.5, 1.4, 2.0) that exceed or match their corresponding $\mu^*$ values (1.2, 1.5, 1.7). When the percentage of model grid cells with calculated heads above the land surface is the metric, $K_{zone1}$ has a large $\sigma$ value (9.6) that exceeds its $\mu^*$ value (8.1). The relatively high values of these $\sigma$ values relative to their respective $\mu^*$ values indicate that the effect of these parameters on the model output is either non-linear or that these parameters interact with other parameters. For example, when $K_{zone1}$ is large, larger $R_{Irrig}$ values result in a lower $RMSE_h$, but when the $K_{zone1}$ is small, large $R_{Irrig}$ values result in a higher $RMSE_h$, indicating an interaction between $K_{zone1}$ and $R_{Irrig}$.

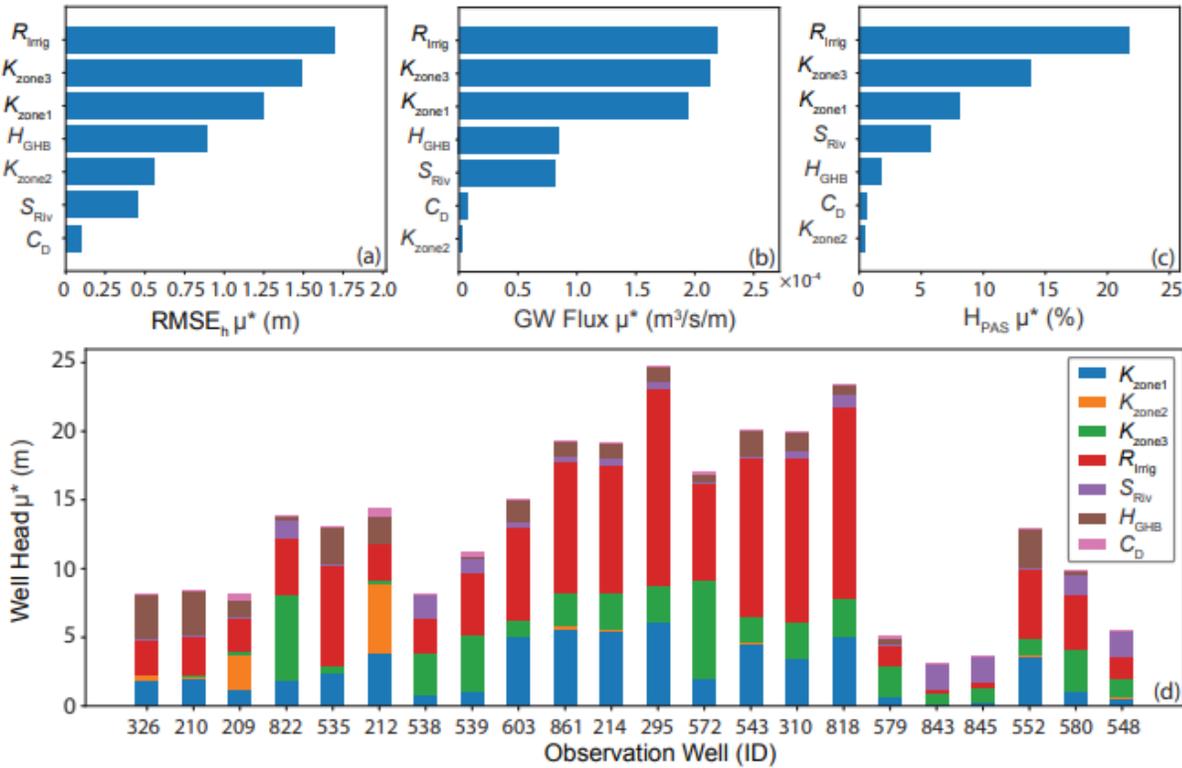

**Figure 6**. Morris Method results using 150 model trajectories. $\mu^*$ is the absolute value of the mean elementary effect. (a) Sensitivity of the $RMSE_h$ to the input parameters. (b) Sensitivity of the groundwater flux rate to the Ticino river to the input parameters. (c) Sensitivity of the percent of calculated heads above the land surface to the model parameters. (d) Sensitivity of the groundwater heads at each observation well to the input parameters. Input parameters include the hydraulic conductivity values of the three aquifer zones ($K_{zone1}$, $K_{zone2}$, $K_{zone3}$), the recharge rate due to irrigation of rice fields ($R_{Irrig}$), the stage of the Ticino River ($S_{Riv}$), the groundwater head at the general head boundaries ($H_{GHB}$), and the conductance of the drain boundaries ($C_D$).

The sensitivity of the calculated heads at each of the observation wells to the model parameters was also assessed (Fig. 6d). The modeled heads are still insensitive to the drain conductance and insensitive to $K_{zone2}$ except for at the two wells located within that zone (i.e., 209 and 212 in Fig. 6d). $K_{zone2}$ is likely not influential due to the relatively small extent of Zone 2. The heads are typically most sensitive to the uncertainty in $K_{zone1}$ and $R_{Irrig}$, followed by $K_{zone3}$, although the relative sensitivity differs according to the location of the well. Additionally, at some wells these parameters are not important factors determining the head value at the well. Wells 843 and 845 are most influenced by the river stage since they are located close to the river. Meanwhile the groundwater level at the general head boundary is an influential parameter at wells 210 and 326 since they are close to the model edge (Fig. 6d). Therefore, while the calculated heads are typically less sensitive to groundwater levels at the general head boundary and river stage, calculated heads in cells near the model edges or rivers are more sensitive to variations in these input parameters.

The results of the Morris analysis are used to help inform model calibration as the method can be used to screen for the parameters to which the model is most and least sensitive (Herman et al., 2013). Since the Morris analysis reveals that the model results are not influenced by the drain conductance parameter, its value is kept constant (100 m$^2$/s) during calibration to reduce the size of the parameter space. While the groundwater level at the general head boundaries and the river stage can influence the model results near the respective boundaries, they are never the most important factors influencing the RMSE$_h$, groundwater flux to the river, or percentage of cells with heads above the land surface. Additionally, calculated and interpolated values based on measured data are available to define the Ticino river stage and the groundwater levels at the general head boundaries during September 2014. Therefore, groundwater levels at the general head boundaries and the river stage were also kept constant, allowing model calibration efforts to focus on determining the values of parameters that are most influential on model results ($K_{zone1}$, $K_{zone3}$, and $R_{Irrig}$).

### 3.3 Calibration using the Negative Log Likelihood

Model simulations were then run varying the values of the hydraulic conductivity in each zone and the aquifer recharge rate due to irrigation activities to estimate the values of these parameters through brute-force optimization. As already explained, the results of this brute-force optimization procedure were then used as starting points for the Nelder-Mead minimization algorithm. The results of the brute-force simulations provide insights into the optimal values of the model parameters and their influence on the model results (Fig. 7). When the hydraulic conductivity of zone 3 is smallest, the RMSE$_h$ can be large depending on the amount of aquifer recharge, and to a lesser extent depending on $K_{zone1}$, with larger recharge rates resulting in larger RMSE$_h$ values (Fig. 7c). When the hydraulic conductivity of zone 3 is greatest, lower recharge rates due to irrigation produce larger RMSE$_h$ values. Meanwhile, intermediate $K_{zone3}$ values can produce either high or low RMSE$_h$ values depending on the recharge rate and $K_{zone1}$ (Fig. 7a-c). When $K_{zone3}$ (the river valley) values are smaller than about $2 \times 10^{-3}$ m/s, the percentage of model cells with heads above the land surface is unrealistic regardless of the values of the other parameters, providing a lower constraint on the possible value of the hydraulic conductivity in the river valley (Fig. 7i). Higher irrigation recharge rates and larger $K$ values produce larger groundwater fluxes to the river (Fig. 7d-f).

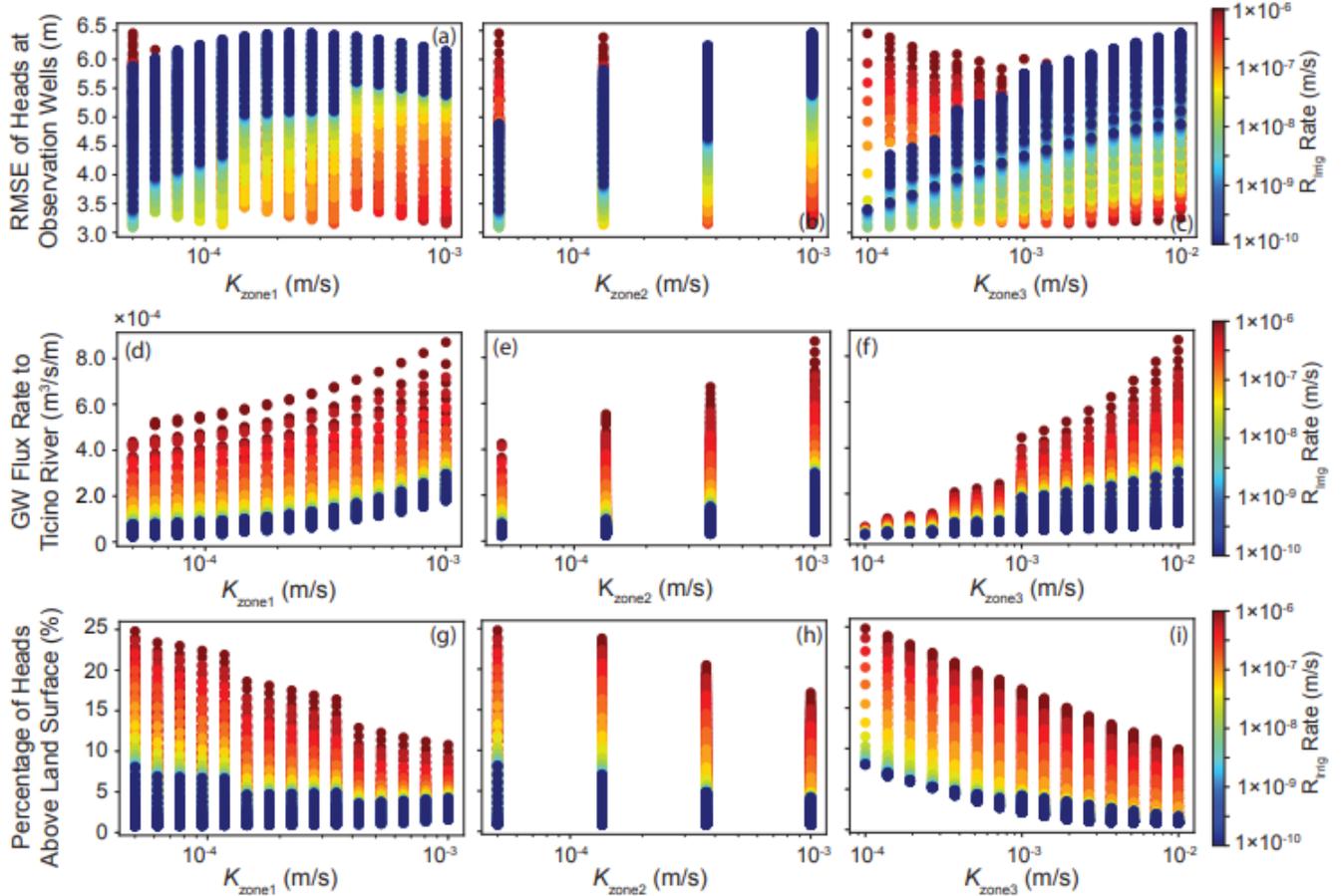

**Figure 7**. Results of the brute-force optimization approach where $K_{zone1}$, $K_{zone2}$, $K_{zone3}$, and $R_{Irrig}$ were all varied over the Cartesian grid constructed within the parameter space. Influence of the recharge rate due to rice field irrigation activities (color of symbol) and the hydraulic conductivity values on the (a-c) $RMSE_h$, (d-f) the groundwater flux rate to the Ticino river, and (g-i) the percentage of cells with heads above the land surface. Input parameters include the hydraulic conductivity values of the three aquifer zones ($K_{zone1}$, $K_{zone2}$, $K_{zone3}$) and the recharge rate due to irrigation of rice fields ($R_{Irrig}$).

Model simulations with the lowest $RMSE_h$ values (<3.1 m) result in calculated heads that exceed the elevation of the land surface in 8-9% of the model cells. When the $RMSE_h$ equals 3.09 m, $H_{PAS}$=8.99%, and 52.3% of fontanili are active with an average discharge rate of 16.9 l/s for $K_{zone1}$=5.0×10$^{-5}$ m/s, $K_{zone2}$=5.0×10$^{-5}$ m/s, $K_{zone3}$=1.0×10$^{-4}$ m/s and $R_{Irrig}$ =7.85×10$^{-9}$ m/s. Meanwhile, the model simulation with the lowest percentage of cells with heads above the land surface ($H_{PAS}$=0.84%) has hydraulic conductivity values in zones 1, 2, and 3 of 5.0×10$^{-5}$, 1.0×10$^{-3}$, 1.0×10$^{-2}$ m/s respectively and an $R_{Irrig}$ of 6.95×10$^{-10}$ m/s, resulting in an $RMSE_h$ of 5.84 m, a groundwater flux rate to the Ticino river of 7.6×10$^{-5}$ m$^3$/s/m, and 19.7% of mapped springs/fontanili active. While this configuration produces a realistic percentage of calculated heads above the land surface, the $RMSE_h$ is too large. In general, model simulations producing the smallest $RMSE_h$ values have too many cells with heads above the land surface (>8%). Meanwhile, model simulations with the lowest percentage of cells above the land surface have large $RMSE_h$ values (>5.8 m). Therefore, a compromise between these quantities had to be introduced to determine the input parameters which most accurately represent the groundwater system, leading to Eq. 7.

The simulation from the brute-force optimization with the lowest joint error metric (NLL) has a $RMSE_h$ of 3.67 m, $H_{PAS}$ of 1.27%, and a groundwater flux rate to the Ticino river of 9.8×10$^{-5}$ m$^3$/s/m when input parameter values of $K_{zone1}$=5.0×10$^{-5}$, $K_{zone2}$=5.0×10$^{-5}$, $K_{zone3}$=5.18×10$^{-3}$ m/s, and $R_{Irrig}$=3.36×10$^{-8}$ m/s are used (see Table 4 run 2267). The NLL value for this result is 64.38 with a $\sigma_h$ of 3.49 m. Fifteen of the parameter sets with the lowest NLL values were then used as starting points for the Nelder-Mead algorithm to refine the optimal parameter values by minimizing the NLL (Table 4). This procedure resulted in a final optimized set of parameter values of $K_{zone1}=8.13\times10^{-5}$, $K_{zone2}=8.13\times10^{-5}$, $K_{zone3}=6.62\times10^{-3}$ m/s, and $R_{Irrig}=5.69\times10^{-8}$ m/s, with a RMSE$_h$ of 3.64 m, $H_{PAS}$ of 1.26%, and a groundwater flux rate to the Ticino river of $1.38\times10^{-4}$ m$^3$/s/m (see Table 4 run 1928). The optimized $\sigma_h$ value is 3.60 m, which is in good agreement with the RMSE$_h$ value of 3.64 m, indicating consistency of the procedure for estimating the noise of the experimental error. After correcting the $R_{Irrig}$ value for excess recharge applied to 0.9% of the rice fields (and removed through drain boundaries), the final $R_{Irrig}$ value in the rice fields is an average of $4.16\times10^{-8}$ m/s.

    The standard deviations of the residual uncertainty of the parameters, obtained as square roots of the diagonal entries of the covariance matrix of the parameters (see Section 2.4), are $3.72\times10^{-6}$, $1.92\times10^{-5}$, $1.60\times10^{-3}$, and $2.14\times10^{-8}$ m/s for $K_{zone1}$, $K_{zone2}$, $K_{zone3}$, and $R_{Irrig}$ respectively. During this simulation 25.0% of the mapped springs and fontanili are active with an average discharge of 79.1 l/s. The map of calculated heads (Fig. 8a) displays a similar pattern to the map of interpolated heads (Fig. 2b). Calculated heads above the land surface where there are no mapped fontanili typically occur in cells near the transition from $K_{zone1}$ to $K_{zone2}$, near cells with mapped springs, and along the upper section of the Ticino where the river is braided (Fig. 8b.). The RMSE$_h$ of 3.64 m indicates an acceptable agreement between calculated and measured heads at the observation wells (Fig. 8d); for example, a groundwater model of a nearby area in Lombardy modelling data measured at 217 wells during 1994 data had an RMSE of 3.47 (De Caro et al., 2020). Calculated heads at 68% of the observation wells have a percent error of 5% or less, and all of the observation wells have percent errors less than 10% (Fig. 8c). Considering the groundwater flow model is a simplified version of the real, complex groundwater system, that fewer wells had available data than other studies (De Caro et al., 2020; Musacchio et al., 2021), and that groundwater levels were measured over the span of a month rather than over the course of a few days, such differences between the modeled and measured heads are quite acceptable and indicate that the groundwater model represents the relevant features of the aquifer system with a suitable degree of reliability.

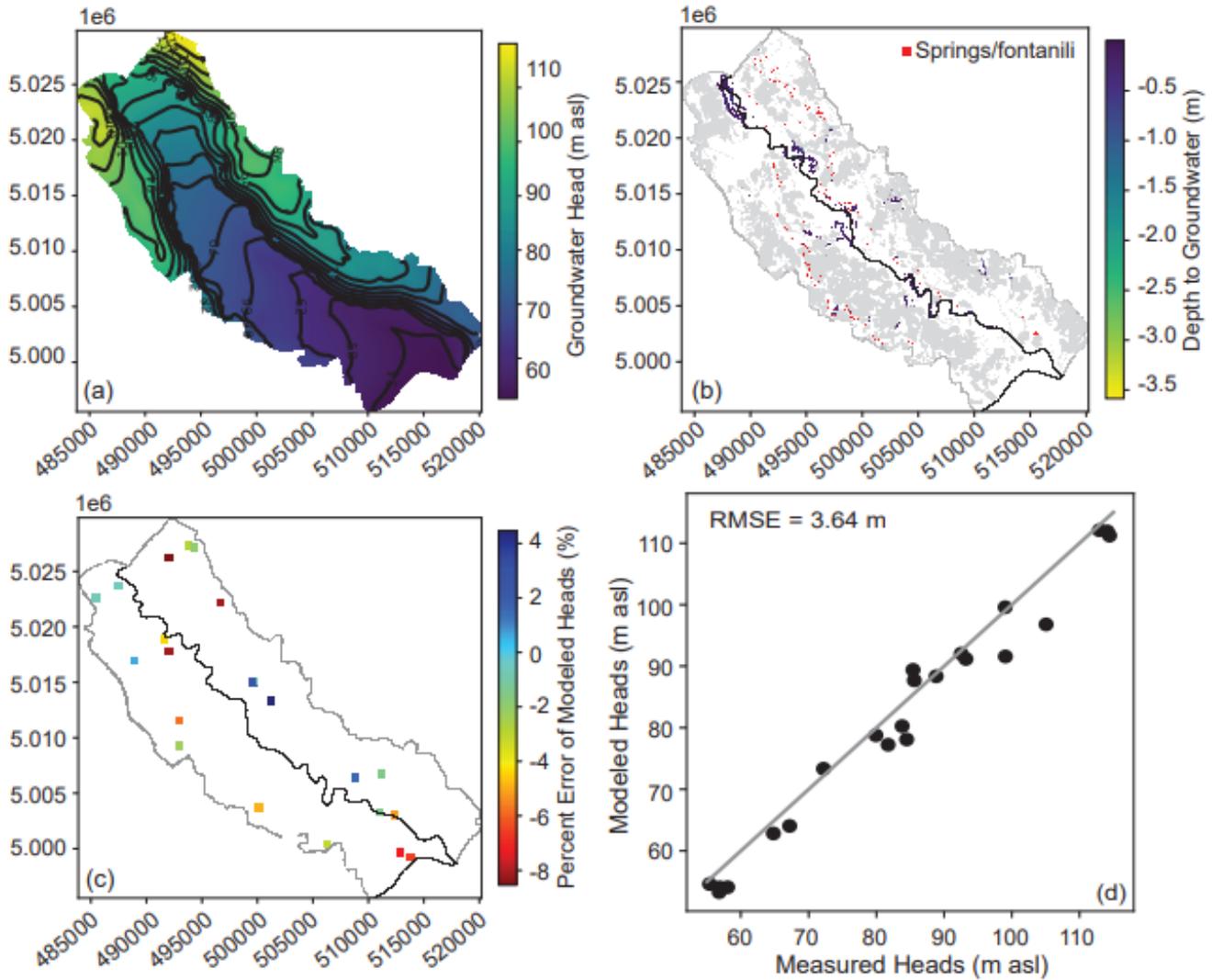

**Figure 8**. Outputs of the optimized groundwater flow model for September 2014. (a) Calculated groundwater heads with black contour lines at 4 m intervals. (b) Map of where calculated groundwater heads are above the land surface elevation. The general head boundary at the domain edges is gray, the constant head boundary representing the rivers is black, the light gray zones are the locations of rice fields, and the locations of the mapped springs and fontanili are shown in red. (c) The percent error ([observed head-modeled head]/observed head*100) between modeled and observed groundwater heads at each observation well. (d) The measured versus modeled groundwater heads along the 1:1 line, with a coefficient of determination ($R^2$) of 0.96.

## 4. Discussion
### 4.1 The Morris Method & Parameter Optimization

The Morris Method was used to determine which model input parameters were the most influential factors affecting multiple model outputs over the range of uncertainty in the input parameters. This sensitivity analysis provides a rigorous approach to assess the contributions of the input parameters to model uncertainty, enabling uncertain parameters that do not significantly affect model outputs to be excluded from further analyses. Previous studies have demonstrated that the Morris method is an ideal sensitivity analysis because failed simulations can easily be removed/substituted and it has a lower computation cost compared to other methods, while still serving as a good proxy for more rigorous analyses such as the Sobol method (Campolongo et al., 2007). A study comparing the computational expense of sensitivity analyses of an energy balance model with similar computational time to our model (1-2 minutes per simulation) found that the Morris method was substantially cheaper than alternative methods for sensitivity analysis, requiring a few tens of runs for Morris versus several hundred to thousands of runs for the other methods (Nguyen & Reiter, 2015). The Morris analysis conducted in this study identified the recharge rate due to irrigation and the hydraulic conductivity of zones 1 and 3 as the most influential input parameters. The input parameters to which the model outputs were insensitive or only locally influential included the river stage, groundwater level at the general head boundary, drain conductance, and the hydraulic conductivity of zone 2. By setting river stage, groundwater level at the general head boundary, and drain conductance constant, the subsequent brute-force optimization was able to be conducted with less computational expense on the remaining uncertain input parameters to which the model outputs are sensitive.

Parameter optimization was conducted by considering simultaneously the RMSE of the heads at the observation wells and the percentage of grid cells with calculated heads above the land surface, which was a proxy for the occurrence of springs and fontanili in the area that are unmapped or differ slightly in location due to model simplifications. By optimizing the NLL (Eq. 7), a set of model input parameter values were selected that are physically consistent and represent the groundwater system more accurately than if only the RMSE was used. The input parameter values that minimize the joint error metric (NLL) are hydraulic conductivity values of $8.13 \times 10^{-5}$, $8.13 \times 10^{-5}$, and $6.62 \times 10^{-3}$ m/s for zones 1, 2, and 3 respectively and an adjusted $R_{Irrig}$ of $4.16 \times 10^{-8}$ m/s. These parameter values produce modeled hydraulic heads with a RMSE of 3.64 m, a groundwater flux rate to the Ticino river of $1.4 \times 10^{-4}$ m$^3$/s/m, 1.26% of cells with calculated heads above the land surface, and 25% of the springs and fontanili active with an average discharge of 79 l/s and maximum discharge of 471 l/s.

The calibrated parameters now follow a gaussian distribution: this means that while we do not rule out the possibility that the calibration is not exact, values further from the calibrated ones are deemed less and less likely. In particular, using the standard deviations of the residual uncertainties, we can provide a reduced range in the uncertainty in the model parameters as the calibrated parameter value plus/minus two times the standard deviation of the residual uncertainty. Using this method, the uncertainties in the values of $K_{zone1}$ and $R_{Irrig}$ have been reduced the most, with the uncertainty in $K_{zone1}$ reduced from 1.5 orders of magnitude to less than a quarter of an order of magnitude and the uncertainty in $R_{Irrig}$ reduced from 4 orders of magnitude to less than 1 order of magnitude (Table 5). Meanwhile, $K_{zone2}$ and $K_{zone3}$ are now both known within less than one order of magnitude (reduced from 1.5 and 2 orders of magnitude respectively (Table 5). While we initially had a poor idea of the recharge rate due to irrigation practices, we now have a much narrower, informative estimate after the implementation of this method. These reduced input parameter ranges can now be used to guide calibration of a transient groundwater flow model in future work or to obtain robust estimates on other quantities of interest output by the model, such as the groundwater flux rate to the river or particle travel times. Robust estimates of these quantities of interest (an expected value along with their range in uncertainty) can be obtained by running the model multiple times for different values of the input parameters within the range of the residual uncertainty.

**4.2 Model Estimated Fluxes**

The estimated fluxes into and out of the groundwater flow model for this optimized set of parameter values are reasonable based on available data and other studies within the region. The groundwater flux rate to the Ticino river estimated by the model results in a net gain in streamflow of about 7.70 m$^3$/s along this segment of the Ticino river. The average streamflow rate at the gauging station in Pavia is 233 m$^3$/s during August and September of 2014. Therefore, the estimated groundwater flux rate into the Ticino river along the modeled reach is roughly 3.3% of the average streamflow in Pavia. Rivers simulated by a groundwater flow model over a ten-year period in the nearby Adda river basin containing portions of the Adda, Tormo, and Serio rivers had an average net gain in streamflow of about 18.5 ± 1.16 m$^3$/s (1.6 ± 0.1 hm$^3$/day; Musacchio et al., 2021) over the approximately 123 km of total river within the model domain. This results in an average gain in streamflow of about 1.5×10$^{-4}$ m$^3$/s/m for rivers within the Adda river basin, as compared to the rate of 1.38×10$^{-4}$ m$^3$/s/m estimated by the groundwater model in the current study. Therefore, the groundwater flux rate estimated by the model in the lower Ticino basin is similar to that estimated within the modeled portion of the Adda basin in Musacchio et al. (2021). Since the hydrogeology of both basins is similar, this helps confirm the accuracy of the groundwater flow rate into the Ticino river estimated by the current study and therefore the values of the model input parameters.

The optimized set of model input parameters also results in a steady-state groundwater flow model that can reproduce the behavior of the springs and fontanili in the study area. The model predicts 25.0% of the mapped springs (risorgive) and fontanili are active in September of 2014, with an average discharge of 79.0 l/s and a maximum discharge rate of 471 l/s. Balestrini et al. (2021) studied about 20 fontanili in the Ticino basin (3 of which were in the study area) that exhibited discharge rates of 1.1-274 l/s with discharge rates varying according to the fontanili location and the time of year. Meanwhile, studied fontanili within the Po Plain in the Turin province exhibited discharge rates ranging from 4-80 l/s (De Luca et al., 2014) and fontanili studied throughout the region of Lombardy had discharge rates ranging from 10 to 1000 l/s (0.01-1.0 m$^3$/s) or more (Fumagalli et al., 2017). Therefore, estimated model discharge rates from springs and fontanili occur within the measured range. Many of the mapped springs and fontanili are not actively discharging water within the model. However, looking at the map of where calculated groundwater heads are above the land surface and comparing it with the fontanili and spring locations (Fig. 8b), it is apparent that many of the cells with calculated heads above the land surface are near locations of mapped springs and fontanili, but not within the same cells. Since these springs are known to occur where there are steep land surfaces intersecting locations where the hydraulic permeability changes abruptly, the springs may occur at slightly different locations in the model than observed due to model simplifications of the hydraulic conductivity values and differences in land surface elevations due to the coarser resolution of the resampled DTM. Therefore, future studies could make the whole area around the mapped springs and fontanili drain cells to better capture their modeled locations resulting from the simplified conceptual model of the surficial aquifer.

The steady-state groundwater model also provides an estimate of the recharge rate due to rice irrigation that agrees with rates determined in previous studies. The model estimates an adjusted recharge rate from rice field irrigation and canal leakage of 4.16×10$^{-8}$ m/s. As mentioned in the introduction, the practice of flood irrigation maintains ponded water depths in rice fields of 5 -10 cm during the growing season, for a total applied irrigation depth of 1500 – 3000 m (Cesari de Maria et al., 2016; Cesari de Maria et al., 2017), resulting in infiltration rates of 2.4 to 15.3 mm/day (Facchi et al., 2018) and up to 10 to 40 mm/day in some fields (Cesari de Maria et al., 2017). This equates to a range in infiltration rates of approximately 2.8×10$^{-8}$ to 4.6×10$^{-7}$ m/s, which encompasses the recharge rate due to rice field irrigation estimated in the current study, further validating the estimated model input parameters. Meanwhile, the estimated recharge rate from precipitation ranges from 2.78×10$^{-9}$ to 9.73×10$^{-9}$ m/s depending on the sub-basin with an average area weighted rate of 5.72×10$^{-9}$ m/s during the study period. This indicates that 81-94% (mean of 87.9%) of the aquifer recharge during this period is due to irrigation activities. A study in the nearby Oglio river basin using a δ$^2$H-Cl/Br mixing model estimated that 55-88% of aquifer recharge is due to irrigation (Rotiroti et al., 2019). The dominant crop in the Oglio basin study area was corn, whereas in the lower Ticino basin the dominant crop is rice (29% of the land surface; Regione Lombardia, 2019). Since flood irrigation is used within the rice fields, it is reasonable that the lower Ticino basin experiences recharge rates due to irrigation on the upper end of the range of those estimated in the Oglio basin where corn cultivation is dominant. Overall, the model results confirm that, during the modeled period, irrigation of rice fields is a major source of recharge to the surficial aquifer in the lower Ticino basin and provide an estimate of the recharge rate of the surficial aquifer due to flood irrigation.

**4.3 Implications, Limitations & Future Work**

The current groundwater flow model provides estimated average values for the hydraulic conductivity of the surficial aquifer in the lower Ticino basin and provides estimates of the groundwater flux rate to the Ticino river and the aquifer recharge rate due to irrigation within rice paddy fields and leakage from unlined irrigation canals and ditches during the end of the irrigation season (August/September 2014). The estimated hydraulic conductivity values and recharge rate due to irrigation (along with their residual uncertainties) can be used in the construction of a future transient groundwater flow model, reducing the uncertainty in the model parameters. Additionally, the estimated groundwater flux rates into the Ticino river and the estimated recharge rates due to irrigation activities provide insights into their roles within the river basin. During the modeled period, the groundwater flux accounted for about 3.3% of the total average streamflow in the Ticino river at the Pavia gauging station, while the irrigation waters applied to rice fields and leaking from canals accounted for about 91% of aquifer recharge. The magnitude of the estimated recharge rate due to irrigation activities highlights the significant influence that flood irrigation practices can have on both groundwater levels and chemistry, as demonstrated in other basins in the Po Plain. For example, recharge from flood irrigation practices in rice fields is known to raise groundwater levels in surficial aquifers by up to 4 m and dilute nitrate concentrations in groundwater in locations when irrigation water is sourced from water containing low nitrate concentrations (Rotiroti et al., 2019). Therefore, any management decisions aimed at making irrigation practices more efficient, or changes in the abundance of rice paddy fields, will affect aquifer recharge and consequently groundwater levels in the basin. By lowering groundwater levels through reduced artificial recharge, the amount of water available to be pumped from shallow wells in the unconfined aquifer could be affected, negatively affecting local small farmers in the study area. Additionally, lowering of the water table would also reduce the flow to springs and fontanili, potentially causing them to dry up. This would not only destroy the local microsystems that form around these groundwater seeps, but it would also remove an additional water resource used by local farmers. Such decreases in groundwater levels due to less artificial surface recharge will only be exacerbated by climate change, as it is predicted to increase temperatures and decrease precipitation in the region (Lasagna et al., 2020).

In the future we plan to collect additional data to develop a transient groundwater model of this area to explore how changes in irrigation practices, crop type and rotation, and climate change could impact groundwater levels in the basin. Future models could also address some of the current model's limitations, such as using alternative methods for estimating the aquifer recharge due to precipitation, implementing more drain cells near the mapped fontanili to allow for spatial flexibility in where springs and fontanili occur, and incorporating groundwater extractions from municipal and agricultural wells (although many of these wells extract groundwater from lower aquifer units; De Caro et al., 2020). Another limitation is that the resulting parameter rankings obtained from the Morris Method are tightly linked to the assessed model outputs and so under transient model conditions or during the evaluation of travel times this initial global sensitivity ranking might break down (though this limitation is also a concern with other approaches, such as when using PEST). Additional limitations of the current model include the simplification of the hydraulic conductivity values into three zones, uncertainty in the true location of rice fields each year due to the practice of crop rotation, and the implementation of the same recharge rate due to both infiltration of water within rice fields and leakage of irrigation water from irrigation canals/ditches since both rates are currently lumped together in the $R_{Irrig}$ parameter. Future models could incorporate a more complex hydraulic conductivity field, randomly assign crop type to the fields known to cultivate both rice and other crops during different years and apply separate recharge rates to the rice fields and irrigation canals/ditches. Since the results show that the model is most sensitive to the recharge inputs, future work could couple a transient groundwater flow model with a hydrologic model for simulating exchanges between surface and sub-surface flows (e.g., GSFLOW (Markstrom et al., 2008), SWAT-MODFLOW (Bailey et al., 2016), HydroGeoSphere (Brunner & Simmons, 2011), Advanced Terrestrial Simulator (Coon et al., 2020)). However, this would increase the model complexity, introducing additional modelling parameters and requiring more detailed data for model calibration and validation.

## 5. Conclusions

A common problem encountered in groundwater flow modeling is the large uncertainty in the values of multiple input parameters. In this work, multiple error metrics are coupled with the Morris Method for sensitivity analysis and joint error metric optimization to estimate the hydraulic conductivity values and recharge rate due to irrigation activities within the lower Ticino basin. The use of the Morris Method as a global sensitivity analysis provides a rigorous yet only moderately computationally expensive approach to identifying which uncertain input parameters are most influential and which ones only have a minimal effect on the groundwater flow model. Uninfluential parameters can then be set constant during subsequent optimization so that influential parameters can be explored more thoroughly. Additionally, optimization of a joint weighted error metric (Negative Log Likelihood) enables the selection of input parameters values that, besides being physically consistent, produce more realistic model outputs than what are produced by minimizing only a single error metric, since the joint metric can consider multiple model behaviors and determine a set of parameter values that produces a suitable compromise, while also estimating the residual uncertainty in the selected parameter values.

Though the current steady-state groundwater flow model of the lower Ticino basin is based on a relatively simple conceptual model of the surficial aquifer and makes some necessary assumptions due to lack of more detailed experimental/field data, it effectively reproduces observed groundwater heads with a RMSE of 3.64 m and reasonably estimates the groundwater flux rate to the river and recharge rate due to irrigation activities. The model confirms that water from the flood irrigation of rice fields is the dominant source (~88%) of recharge to the surficial aquifer during this period. Furthermore, the model reproduces the occurrence of springs along the fontanili line where steep changes in the land surface coupled with changes in hydraulic conductivity cause water to emerge at the surface. Future research can use these results to develop a transient groundwater flow model in the basin to better understand how the relative importance of recharge sources varies over the year, assess groundwater residence time and transport of solutes released by human activities within the surficial aquifer, and explore how changes in irrigation practices and climate could affect groundwater levels and flow within the surficial aquifer.

**Declaration of competing interest**
The authors declare that they have no known competing financial interests or personal relationships that could have appeared to influence the work reported in this paper.

**Acknowledgements**
This research was supported by Regione Lombardia, POR FESR 2014-2020 - Call HUB Ricerca e Innovazione, Progetto 1139857 CE4WE: Approvvigionamento energetico e gestione della risorsa idrica nell'ottica dell'Economia Circolare (Circular Economy for Water and Energy). A. Reali and S. Manenti were also partially supported by the Italian Ministry of Education, University and Research (MIUR) through the PRIN project XFAST-SIMS (No. 20173C478N). L. Tamellini has been supported by the PRIN 2017 project 201752HKH8 "Numerical Analysis for Full and Reduced Order Methods for the efficient and accurate solution of complex systems governed by Partial Differential Equations (NA-FROM-PDEs)". The authors are grateful to Regione Lombardia and the Agenzia Regionale per la Protezione dell'Ambiente (ARPA) for access to monitoring data. The authors thank: Prof. Giorgio Pilla (University of Pavia) for spring/fontanili data, data on the depth to the aquifer base, and for the aquifer conceptualization; Dr. Maurizio Gorla (Gruppo CAP Milano) and Fasani Maurizio (Studio Trilobite, Via San L. Beccari n. 2 - Gropello Cairoli) for stratigraphy and borehole log data.

Rotiroti, M., Bonomi, T., Sacchi, E., McArthur, J. M., Stefania, G. A., Zanotti, C., Taviani, S., Patelli, M., Nava, V., Soler, V., Fumagalli, L., & Leoni, B. (2019). The effects of irrigation on groundwater quality and quantity in a human-modified hydro-system: The Oglio River basin, Po Plain, Northern Italy. *Science of the Total Environment*, 672, 342–356. https://doi.org/10.1016/j.scitotenv.2019.03.427

Ruano, M. V., Ribes, J., Seco, A., & Ferrer, J. (2012). An improved sampling strategy based on trajectory design for application of the Morris method to systems with many input factors. *Environmental Modelling and Software*, 37, 103–109. https://doi.org/10.1016/j.envsoft.2012.03.008

Saleh F., A. Ducharne, N. Flipo, L. Oudin, E. Ledoux. (2013). Impact of riverbed morphology on discharge and water levels simulated by a 1D De Saint-Venant hydraulic model at regional scale. *Journal of Hydrology*, 476, 169-177. https://doi.org/10.1016/j.jhydrol.2012.10.027

Vassena, C., Rienzner, M., Ponzini, G., Giudici, M., Gandolfi, C., Durante, C., & Agostani, D. (2012). Modeling water resources of a highly irrigated alluvial plain (Italy): Calibrating soil and groundwater models. *Hydrogeology Journal*, 20(3), 449–467. https://doi.org/10.1007/s10040-011-0822-2

Zampieri, M., Ceglar, A., Manfron, G., Toreti, A., Duveiller, G., Romani, M., Rocca, C., Scoccimarro, E., Podrascanin, Z., & Djurdjevic, V. (2019). Adaptation and sustainability of water management for rice agriculture in temperate regions: The Italian case-study. *Land Degradation and Development*, 30(17), 2033–2047. https://doi.org/10.1002/ldr.3402

# TABLES

**Table 1**. Land surface area, impervious area, percent of the land surface that is impervious, and the estimated curve numbers (CN) for each hydrologic sub-basin.

| Basin | Area [km$^2$] | Impervious Area [km$^2$] | % Impervious | CN |
|---|---|---|---|---|
| **i: Abbiategrasso** | 70.84 | 8.46 | 11.95 | 73.50 |
| **ii: Vigevano** | 51.24 | 15.46 | 30.17 | 76.85 |
| **iii: Motta Visconti** | 68.64 | 8.23 | 11.99 | 74.16 |
| **iv: Scavizzolo** | 57.85 | 4.69 | 8.11 | 62.27 |
| **v: Pavia** | 58.93 | 10.96 | 18.61 | 77.93 |
| **vi: Mangialoca** | 68.65 | 2.83 | 4.12 | 63.63 |
| **vii: Roggia Vernavola** | 47.42 | 9.07 | 19.12 | 77.49 |
| **viii: Gravellone** | 77.99 | 10.07 | 12.91 | 71.89 |
| **Total** | 501.57 | 69.05 | 13.92 | 71.93 |

**Table 2**. Model input parameters and values tested using the Morris Analysis. Each input parameter is sampled at 6 equispaced values across its range of uncertainty. Input parameters with uncertainties larger than an order of magnitude ($K$, $R_{Irrig}$, $C_D$) are sampled in log space. Input parameters include the hydraulic conductivity values of the three aquifer zones ($K_{zone1}$, $K_{zone2}$, $K_{zone3}$), the recharge rate due to irrigation of rice fields ($R_{Irrig}$), the stage of the Ticino River ($S_{Riv}$), the groundwater head at the general head boundaries ($H_{GHB}$), and the conductance of the drain boundaries ($C_D$).

| $K_{zone1}$ (m/s) | $K_{zone2}$ (m/s) | $K_{zone3}$ (m/s) | $R_{Irrig}$ (m/s) | $S_{Riv}$ (m) | $H_{GHB}$ (m) | $C_D$ (m²/s) |
|---|---|---|---|---|---|---|
| 5.00E-05 | 5.00E-05 | 1.00E-04 | 1.00E-10 | -1.0 | -2.0 | 0.10 |
| 9.10E-05 | 9.10E-05 | 2.51E-04 | 6.31E-10 | -0.6 | -1.2 | 0.40 |
| 1.66E-04 | 1.66E-04 | 6.31E-04 | 3.98E-09 | -0.2 | -0.4 | 0.63 |
| 3.02E-04 | 3.02E-04 | 1.59E-03 | 2.51E-08 | 0.2 | 0.4 | 6.31 |
| 5.50E-04 | 5.50E-04 | 3.98E-03 | 1.58E-07 | 0.6 | 1.2 | 25.12 |
| 1.00E-03 | 1.00E-03 | 1.00E-02 | 1.00E-06 | 1.0 | 2.0 | 100.00 |

**Table 3**. The parameter values tested during brute-force optimization, resulting in 6300 parameter combinations. Parameters include the hydraulic conductivity values of the three aquifer zones ($K_{zone1}$, $K_{zone2}$, $K_{zone3}$) and the recharge rate due to irrigation of rice fields ($R_{Irrig}$).

| | Model Parameters | | | |
|---|---|---|---|---|
| | $K_{zone1}$ (m/s) | $K_{zone2}$ (m/s) | $K_{zone3}$ (m/s) | $R_{Irrig}$ (m/s) |
| | 1.00E-03 | 1.00E-03 | 1.00E-02 | 1.00E-06 |
| | 8.07E-04 | 3.68E-04 | 7.20E-03 | 6.16E-07 |
| | 6.52E-04 | 1.36E-04 | 5.18E-03 | 3.79E-07 |
| | 5.26E-04 | 5.00E-05 | 3.73E-03 | 2.34E-07 |
| | 4.25E-04 | - | 2.68E-03 | 1.44E-07 |
| | 3.43E-04 | - | 1.93E-03 | 8.86E-08 |
| | 2.77E-04 | - | 1.39E-03 | 5.46E-08 |
| | 2.24E-04 | - | 1.00E-03 | 3.36E-08 |
| | 1.81E-04 | - | 7.20E-04 | 2.07E-08 |
| | 1.46E-04 | - | 5.18E-04 | 1.27E-08 |
| Values | 1.18E-04 | - | 3.73E-04 | 7.85E-09 |
| | 9.50E-05 | - | 2.68E-04 | 4.83E-09 |
| | 7.67E-05 | - | 1.93E-04 | 2.98E-09 |
| | 6.19E-05 | - | 1.39E-04 | 1.83E-09 |
| | 5.00E-05 | - | 1.00E-04 | 1.13E-09 |
| | - | - | - | 6.95E-10 |
| | - | - | - | 4.28E-10 |
| | - | - | - | 2.64E-10 |
| | - | - | - | 1.62E-10 |
| | - | - | - | 1.00E-10 |

Table 4. Nelder-Mead optimization using the brute-force results with the lowest NLL values as starting points for the optimization algorithm. Parameters include the hydraulic conductivity values of the three aquifer zones ($K_{zone1}$, $K_{zone2}$, $K_{zone3}$) and the recharge rate due to irrigation of rice fields ($R_{Irrig}$). $K_{zone2}$ was constrained such that it had to be greater than or equal to $K_{zone1}$. The algorithm typically converged to minima where $K_{zone2}$ equaled $K_{zone1}$.

| NLL Brute Force Results (Initial Inputs) | | | | | Optimization Results | | | | | |
|---|---|---|---|---|---|---|---|---|---|---|
| Run | $K_{zone1}$ (m/s) | $K_{zone2}$ (m/s) | $K_{zone3}$ (m/s) | $R_{Irrig}$ (m/s) | NLL | $K_{zone1-2}$ (m/s) | $K_{zone3}$ (m/s) | $R_{Irrig}$ (m/s) | NLL | $H_{PAS}$ (%) | $RMSE_h$ (m) |
| 2267 | 5.00e-5 | 5.00e-5 | 5.18e-3 | 3.36e-8 | **59.985** | 5.72e-5 | 4.89e-3 | 3.35e-8 | 59.816 | 1.218 | 3.663 |
| 2613 | 5.00e-5 | 5.00e-5 | 3.73e-3 | 2.07e-8 | 60.009 | 6.90e-5 | 5.69e-3 | 4.06e-8 | 59.827 | 1.186 | 3.675 |
| 1588 | 1.18e-4 | 1.36e-4 | 1.00e-2 | 8.86e-8 | 60.044 | 1.19e-4 | 1.00e-2 | 9.22e-8 | 59.829 | 1.246 | 3.655 |
| 2236 | 5.00e-5 | 5.00e-5 | 7.20e-3 | 3.36e-8 | 60.106 | 6.09E-5 | 4.82e-3 | 3.64e-8 | 59.807 | 1.254 | 3.648 |
| 2205 | 5.00e-5 | 5.00e-5 | 1.00e-2 | 3.36e-8 | 60.300 | 1.18e-4 | 1.00e-2 | 9.12e-8 | 59.828 | 1.244 | 3.656 |
| 2582 | 5.00e-5 | 5.00e-5 | 5.18e-3 | 2.07e-8 | 60.302 | 4.89e-5 | 4.01e-3 | 2.44e-8 | 59.872 | 1.206 | 3.676 |
| 2644 | 5.00e-5 | 5.00e-5 | 2.68e-3 | 2.07e-8 | 60.523 | 7.04e-5 | 6.19e-3 | 4.73e-8 | 59.783 | 1.224 | 3.655 |
| 1619 | 1.18e-4 | 1.36e-4 | 7.20e-3 | 8.86e-8 | 60.586 | 1.11e-4 | 8.77e-3 | 8.15e-8 | 59.812 | 1.239 | 3.655 |
| 1931 | 9.50e-5 | 1.36e-4 | 7.20e-3 | 5.46e-8 | 60.601 | 8.54e-5 | 6.90e-3 | 6.04e-8 | 59.772 | 1.254 | 3.643 |
| 2551 | 5.00e-5 | 5.00e-5 | 7.20e-3 | 2.07e-8 | 60.603 | 5.08e-5 | 4.14e-3 | 2.62e-8 | 59.858 | 1.217 | 3.670 |
| 2298 | 5.00e-5 | 5.00e-5 | 3.73e-3 | 3.36e-8 | 60.649 | 5.26e-5 | 4.29e-3 | 2.79e-8 | 59.846 | 1.218 | 3.668 |
| 1928 | 7.7e-5 | 1.36e-4 | 7.20e-3 | 5.46e-8 | 60.672 | 8.13e-5 | 6.62e-3 | 5.69e-8 | **59.771** | 1.257 | 3.642 |
| 1585 | 9.50e-5 | 1.36e-4 | 1.00e-2 | 8.85e-8 | 60.770 | 1.06e-4 | 1.00e-2 | 8.29e-8 | 59.855 | 1.236 | 3.663 |
| 1897 | 7.7e-5 | 1.36e-4 | 1.00e-2 | 5.46e-8 | 60.857 | 1.18e-4 | 1.00e-2 | 9.14e-8 | 59.829 | 1.245 | 3.656 |
| 2520 | 5.00e-5 | 5.00e-5 | 1.00e-2 | 2.07e-8 | 60.865 | 1.21e-4 | 1.00e-2 | 9.17e-8 | 59.835 | 1.237 | 3.659 |

Table 5. Initial and final ranges of the model input parameter values. The final parameter ranges are calculated as two times the standard deviation of the residual uncertainty of the parameters. The true values of the parameters are more likely near the center of the final parameter ranges since the calibrated parameters are assumed to follow a gaussian distribution. Parameters include the hydraulic conductivity values of the three aquifer zones ($K_{zone1}$, $K_{zone2}$, $K_{zone3}$) and the recharge rate due to irrigation of rice fields ($R_{Irrig}$).

|  | $K_{zone1}$ (m/s) | $K_{zone2}$ (m/s) | $K_{zone3}$ (m/s) | $R_{Irrig}$ (m/s) |
|---|---|---|---|---|
| Initial Parameter Uncertain Range | 5.0e-5 – 1.0e-3 | 5.0e-5 – 1.00e-3 | 1.0e-4 – 1.0e-2 | 1.0e-10 – 1.0e-6 |
| Final Parameter Range | 7.39e-5 – 8.87e-5 | 4.29e-5 – 1.20e-4 | 3.42e-3 – 9.82e-3 | 1.40e-8 – 9.98e-8 |